\nonstopmode

\newcommand{\ie}{i.e.{}}
\newcommand{\eg}{e.g.{}}
\newcommand{\re}{{\mathrm{Re}\,}}
\newcommand{\im}{{\mathrm{Im}\,}}
\newcommand{\eV}{\U{eV}}
\newcommand{\mul}{\cdot}
\newcommand{\Hartree}{\U{E_{\mathit h}}}
\newcommand{\angstrom}{\U{\hbox{\AA}}}
\newcommand{\Cal}[1]{{\cal #1}}
\newcommand{\etal}{\textit{et al.}}
\newcommand{\U}[1]{\,{\rm{#1}}}
\newcommand{\I}[1]{_{\mathrm{#1}}}
\newcommand{\imag}{{\rm i}}
\newcommand{\euler}{\mathrm e}
\newcommand{\Sum}{\sum\limits}
\newcommand{\Int}{\int\limits}
\newcommand{\Lim}{\lim\limits}
\newcommand{\differential}{\>\mathrm d}
\newcommand{\bra}[1]{\left<\right.\!#1\!\left.\right|}
\newcommand{\ket}[1]{\left|\right.\!#1\!\left.\right>}
\newcommand{\cleb}[2]{C(#1 ; #2)}
\newcommand{\threeY}[2]{\Cal Y(#1 ; #2)}
\newcommand{\E}[1]{\times 10^{#1}}
\newcommand{\XUV}{\textsc{xuv}}
\newcommand{\NIR}{\textsc{nir}}
\newcommand{\atopa}[2]{\genfrac{}{}{0pt}{}{#1}{#2}}

\documentclass[pra,aps,twocolumn,showpacs,nopreprintnumbers]{revtex4}
\usepackage{bm,bbm,amsmath,amssymb,MnSymbol,subeqnarray,graphicx}
\usepackage[nativepdf,bookmarks,bookmarksopen,bookmarksnumbered,raiselinks,breaklinks,%
pdftitle={Theory of Auger decay by laser-dressed atoms},%
pdfauthor={Christian Buth, Kenneth J. Schafer},%
pdfsubject={Atomic physics},%
pdfkeywords={quantum optics, XUV, photoabsorption, Auger decay, %
light-matter interaction, krypton, attosecond}]{hyperref}

\begin{document}
\title{Theory of Auger decay by laser-dressed atoms}
\author{Christian Buth}
\thanks{Corresponding author}
\email{christian.buth@web.de}
\author{Kenneth J.~Schafer}
\affiliation{Department of Physics and Astronomy, Louisiana State
University, Baton Rouge, Louisiana~70803, USA}
\date{August 27, 2009}

\begin{abstract}
We devise an \emph{ab initio} formalism for the quantum dynamics of
Auger decay by laser-dressed atoms which are inner-shell ionized by
extreme ultraviolet~(\XUV)~light.
The optical dressing laser is assumed to be sufficiently weak
such that ground-state electrons are neither excited nor ionized by it.
However, the laser has a strong effect on continuum electrons which
we describe in strong-field approximation with Volkov waves.
The \XUV{}~light pulse has a low peak intensity and its interaction is
treated as a one-photon process.
The quantum dynamics of the inner-shell hole creation with subsequent
Auger decay is given by equations of motion~(EOMs).
For this paper, the EOMs are simplified in terms of an essential-states
model which is solved analytically and averaged over
magnetic subshells.
We apply our theory to the $M_{4,5} N_1 N_{2,3}$~Auger decay of a
$3d$~hole in a krypton atom.
The orbitals are approximated by scaled hydrogenic wave
functions.
A single attosecond pulse produces $3d$~vacancies which Auger decay
in the presence of an~$800 \U{nm}$~laser with an intensity of~$10^{13}
\U{W \, cm^{-2}}$.
We compute the Auger electron spectrum and assess the convergence of the
various quantities involved.
\end{abstract}

%
%
%

\preprint{arXiv:0905.3756}
\maketitle

\renewcommand{\onlinecite}[1]{\cite{#1}}

\section{Introduction}

The inner-shell ionization of atoms leads to a fascinating array
of many-electron effects.
Such vacancies decay on an ultrafast time scale by fluorescence or
electronic decay.
Electronic decay refers to Auger decay~\cite{Auger:-23} and
its special variant, Coster-Kronig decay~\cite{Coster:NT-35}.
For low energy transitions in, \eg, light elements or high-lying
inner shells in heavier elements, Auger decay is
the dominant relaxation process~\cite{Als-Nielsen:EM-01,Thompson:XR-01}.
For deep inner-shell holes, the ion relaxes in terms of cascades
of fluorescence and electronic decay processes.
All atomic electrons are involved either directly in the photoionization
with subsequent electronic or fluorescence decay or indirectly due to a
rearrangement of the atomic electrons in the presence of newly
formed holes, so-called core relaxation~\cite{Saha:NT-90}.
Furthermore, at photon energies near the ionization threshold,
the outgoing photoelectrons are slow and may interact significantly
with the ionic remnant.
If the subsequent electronic decay process is fast, outgoing photo-
and Auger electrons even repel each other appreciably.
This phenomenon is dubbed post-collision
interaction~\cite{Aberg:EP-82,Saha:NT-90,Gelmukhanov:RX-99,Armen:RA-00}.
Generally, we will refer to electronic decay and Auger decay and
will not explicitly distinguish Coster-Kronig decay.
They are fundamental processes that are pure manifestations of
electron correlations.
In this way, electronic decay processes are ideal for an investigation
with the methods of attosecond
science~\cite{Agostini:PA-04,Scrinzi:AP-06,Bucksbaum:FA-07,%
Niikura:AA-07,Krausz:AP-09} which aim to measure and control the motion
of electrons on their natural time scale, which is the
attosecond.

Attosecond light was first used to measure the Auger decay time of $3d$~vacancies
($M$~shell) in krypton atoms~\cite{Drescher:TR-02,Drescher:AS-05}.
Such vacancies undergo $M_{4,5} N_1 N_{2,3}$~Auger decay which
has been studied experimentally in the frequency
domain~\cite{Aksela:CE-84,Carlson:AD-89,Jurvansuu:IL-01,Schmidtke:AD-01}.
The time-domain study of Auger decay represents a seminal experiment in
several ways.
On the one hand, it demonstrates the power of the newly
created attosecond methodology by comparing its results with existing data.
On the other hand, the study of transient electron motion with
attosecond science has so far been restricted to mostly
one-electron processes, \eg,
Refs.~\onlinecite{Johnsson:AC-07,Mauritsson:CE-08,Krausz:AP-09}
and references therein.
However, the most profound goal of attosecond science remains the study
of electron correlations.

Clearly, controlling a process on an attosecond time scale
requires extreme ultraviolet~(\XUV{})~light for its short cycle period.
Naturally, \XUV{}~light targets inner shells for which the photoabsorption
cross section is highest at these wavelengths.
The degree of control over inner-shell electron motion is potentially
limited compared with valence electrons because of weak present-day
attosecond light sources and the fact that postgeneration
pulse shaping capabilities in the \XUV{} and x-ray domains are
severely limited, \eg, only an amplitude shaping can be accomplished
so far with electromagnetically induced transparency for
x~rays~\cite{Buth:ET-07}.
However, the attosecond light can be augmented by an additional optical laser,
a so-called two-color problem.
For moderate intensities, the influence of the optical laser on electrons
in the atomic ground state and hence the two-electron interaction among
them can be neglected, \eg, noble gases can sustain very high electric
fields before ionizing.
For all elements, the impact of the optical laser on inner-shell electrons
is negligible.
For the optical laser to impact inner-shell electrons, its intensity
needs to be so high that it would valence ionize the atom.
Yet it can be used to structure the continuum, \ie,
have an influence on liberated electrons, and in that way enable
profound control over electronic processes~\cite{Buth:ET-07,Krausz:AP-09}.
A seminal experiment along the above-mentioned lines---but on a much
slower (picosecond) time scale---is ultrafast laser control, using
coherent excitation with short laser pulses, of the energy and proximity
of Rydberg electrons in an atom by Pisharody and Jones~\cite{Pisharody:PT-04}.
To demonstrate their ability to control electron dynamics,
they showed that the autoionization of the doubly-excited barium atoms is
due to electron-electron collisions instead of a slow transfer of energy.
Another notable work is the theoretical investigation of an
\XUV{}~pump-probe scheme for the study of the simultaneous two-electron
emission in helium by Hu and Collins~\cite{Hu:AP-06}.

The time-domain measurement of the Auger decay of the $3d$~vacancy in
krypton atoms~\cite{Drescher:TR-02} has been approached theoretically in
two different ways~\cite{Yakovlev:TR-02,Smirnova:QC-03,Yakovlev:ES-03}.
First, Yakovlev and Scrinzi~\cite{Yakovlev:TR-02} devised a model to
support the experimental study~\cite{Drescher:TR-02}.
They made a simple rate equation ansatz to represent the $3d$~hole
population.
This population was used together with the Auger electron wave function
to replace the transition dipole matrix element in
a formula to determine the laser-streaked photoelectron spectra of
\XUV{}-ionized atoms~\cite{Kitzler:AX-02}.
Second, Smirnova~\etal~\cite{Smirnova:QC-03} revisited the question of
Auger decay by weakly laser-dressed atoms in terms of a fully coherent
system of equations of motion~(EOMs) formulation for an essential-states
model~\cite{Fedorov:AF-97} using Hartree products.
It is constructed to describe the Auger electron spectrum as simply
as possible decoupling the EOMs in terms of a parametric decay width from
Weisskopf-Wigner theory~\cite{Weisskopf:-30,Sakurai:MQM-94,Merzbacher:QM-98}
and solving the resulting system of equations analytically.
Based on the ideas in
Refs.~\onlinecite{Yakovlev:TR-02,Smirnova:QC-03,Yakovlev:ES-03}
Zhao and Lin~\cite{Zhao:TL-05} and
Wickenhauser~\etal{}~\cite{Wickenhauser:TR-05} studied Fano resonances.
Kazansky and Kabachnik~\cite{Kazansky:NT-05,Kazansky:TD-07} developed an
\emph{ab initio} theory for the solution of the time-dependent Schr\"odinger
equation for photoionization of inner atomic shells in terms of a Fano-Feshbach
formalism with short pulses that takes into account near-threshold effects.
Finally, Smirnova~\etal~\cite{Smirnova:EC-05} applied the theory
in Ref.~\onlinecite{Smirnova:QC-03} to devise a scheme to use
electron correlations to make attosecond measurements without
attosecond pulses.

Our study goes beyond previous
work~\cite{Yakovlev:TR-02,Smirnova:QC-03,Yakovlev:ES-03}
and overcomes many of its restrictions.
We develop a nonrelativistic multideterminantial \emph{ab initio} formalism
for the interaction of two-color light with atoms.
We set out from the Hartree-Fock-Slater~(HFS) approximation for the atomic
orbitals.
Such mean-field orbitals are typically a good representation to describe
Auger decay~\cite{Aberg:EP-82}.
Using general spin-singlet configuration-state functions, we
derive EOMs.
We treat the interaction with light semiclassically because, in contrast to a
one-electron quantum electrodynamic formalism~\cite{Buth:TX-07,Buth:AR-08},
band width and pulse duration are treated more easily.
Naturally, these play a decisive role in attosecond science.
The general EOMs are subsequently simplified to an essential-states
model~\cite{Fedorov:AF-97} and the equations are solved analytically in
this special case considering, in contrast to the work in
Ref.~\onlinecite{Smirnova:QC-03}, also the laser dressing of the
photoelectrons.
Furthermore, the laser dressing is, in our case, not required to be weak,
yet the intensity should remain below the excitation and ionization threshold
of atomic ground-state electrons.
We make a model for the atomic electronic structure
in terms of scaled hydrogenic wave
functions~\cite{Yakovlev:TR-02,Yakovlev:ES-03}.
Our formalism is a basis for the study of more complex
situations in Auger decay and its control.
In forthcoming papers~\cite{Buth:AD-09}, we will investigate the
interference between Auger electrons from a twin \XUV{}~attosecond pulse.
Further, one can examine what new avenues for the control of Auger processes
open up when one relaxes the assumption of an essential-states
model to a multichannel treatment.

The paper is structured as follows.
In Sec.~\ref{sec:auger}, we devise EOMs to treat the quantum dynamics of
Auger decay on an \emph{ab initio} level.
Volkov waves are introduced in Sec.~\ref{sec:laserdre} to describe the
laser dressing.
The EOMs are solved analytically for an essential-states
model in Sec.~\ref{sec:essstmo}.
We devote Sec.~\ref{sec:elstructmodel} to the determination of the dipole
and two-electron matrix elements for our formalism.
We then apply our theory to the laser-dressed Auger decay of krypton~$3d$
vacancies;
computational details are given in Sec.~\ref{sec:compdet} and
results are presented in Sec.~\ref{sec:results}.
Conclusions are drawn in Sec.~\ref{sec:conclusion}.

Our equations are formulated in atomic units~\cite{Szabo:MQC-89}.
The Bohr radius~$1 \U{bohr} = 1 \, a_0$ is the unit of length
and $1 \, t_0$ represents the unit of time.
The unit of energy is~$1 \U{hartree} = 1 \Hartree$.
Generally, we use the indices~$h$, $i$, $j$, $m$, \ldots{}
to denote occupied orbitals, $a$, $b$, $c$, $d$, \ldots{} for unoccupied
orbitals, and $p$, $q$, $r$, $s$, \ldots{} for orbitals which may be
occupied or unoccupied.

\section{Quantum dynamics of photoionization and Auger decay}
\label{sec:auger}

This section forms the core of our theory.
In Sec.~\ref{sec:problem}, we describe the schematic
of \XUV{}~photoionization with subsequent Auger decay.
The quantum mechanical foundation is laid out in terms of an \emph{ab initio}
description in Secs.~\ref{sec:elstruct}, \ref{sec:atomXUV}, and
\ref{sec:excitedstates}, where we introduce the Hamiltonian and the
states involved.
The full nonrelativistic formalism is simplified using an
approximate Hamiltonian in Sec.~\ref{sec:modelHam} which comprises
only those two-electron integrals which are essential for Auger decay.
We correct for our omissions by adjusting the energies of the involved
states appropriately.
Finally, we use the time-dependent Schr\"odinger equation with
the approximate Hamiltonian to formulate EOMs in
Sec.~\ref{sec:EOM} for the quantum dynamics of
\XUV{}~absorption and Auger decay.

\subsection{Schematic of the processes}
\label{sec:problem}

\begin{figure}
  \begin{center}
    \includegraphics[clip,width=8cm]{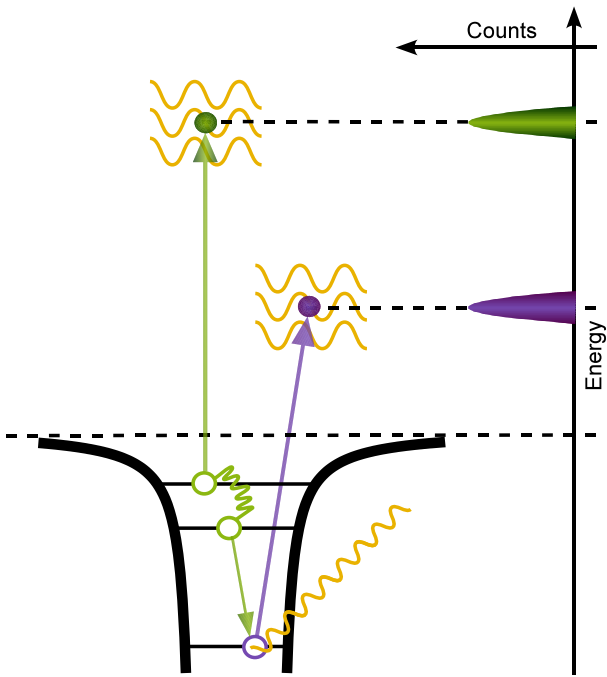}
    \caption{(Color online) Schematic of laser-dressed \XUV{}~photoionization
             of an inner-shell electron with a subsequent Auger decay.
             Only participating electrons are drawn.
             Photoelectron and Auger electron lines are influenced by a
             dressing laser and are observed in an electron spectrometer.
             Similar to Fig.~1a in Ref.~\onlinecite{Drescher:TR-02}.}
    \label{fig:Auger}
  \end{center}
\end{figure}

Auger decay can be treated theoretically in various degrees of
sophistication~\cite{Aberg:EP-82,Aberg:UT-92}.
We focus on a description in terms of a single isolated resonance
(no decay cascades).
In the language of a full scattering process, Auger decay is
a resonance in the double photoionization cross
section~\cite{Aberg:EP-82,Aberg:UT-92}.
The scattering of an \XUV{}~photon~$\gamma$ off an atom~A
leads to the formation of a dication~$\mathrm{A}^{++}$
and the emission of two electrons~$e\I{P}^-$ and $e\I{A}^-$,
\begin{equation}
  \label{eq:AugerScattering}
  \mathrm{A} + \gamma \longrightarrow \mathrm{A}^{++} + e\I{P}^-
    + e\I{A}^- \; .
\end{equation}
Here, we ignore electron correlations among the ground-state electrons
and in the cation and the dication.
We also ignore the interaction between the outgoing electrons and the
remaining ground-state electrons and
the repulsion between the outgoing electrons, so-called
post-collision interactions~\cite{Saha:NT-90,Armen:RA-00}.

\begin{figure}
  \begin{center}
    \includegraphics[clip,width=\hsize]{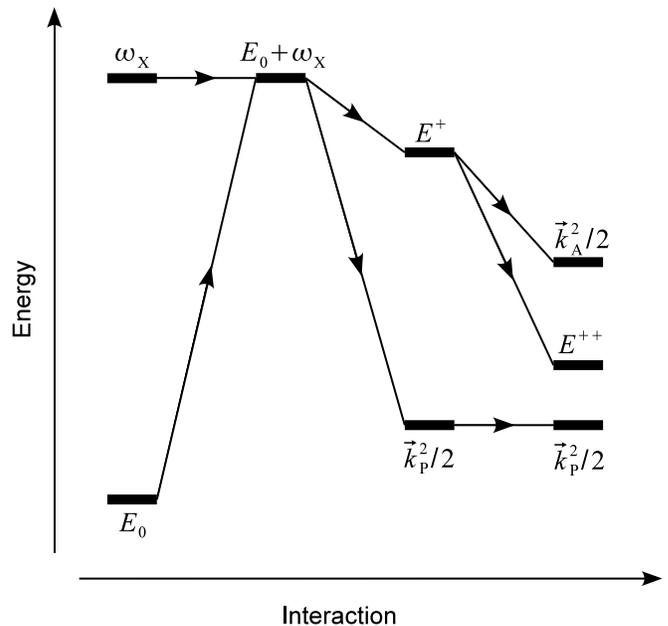}
    \caption{Energy level scheme of the production of an
             inner-shell hole by \XUV{}~photon absorption and its
             subsequent Auger decay.
             The \XUV{}~photon energy is~$\omega\I{X}$, the atomic
             ground-state energy is~$E_0$, the singly-ionized state
             has an energy~$E^+$, and the doubly-ionized state
             energy is~$E^{++}$.
             The photoelectron has a nominal kinetic energy of~$\vec k^2\I{P}/2
             = \Omega\I{P}$ and the Auger electron of~$\vec k^2\I{A}/2
             = \Omega\I{A}$.}
    \label{fig:Kr_levels}
  \end{center}
\end{figure}

Assuming intermediate singly-ionized resonance states, we
break up Eq.~(\ref{eq:AugerScattering}) into two separate processes:
the photoionization of an atomic inner shell
with subsequent Auger decay~\cite{Aberg:EP-82,Aberg:UT-92} which is depicted
in Fig.~\ref{fig:Auger}.
To begin with, let us disregard laser dressing.
A level scheme of the states participating in the photoionization
and subsequent Auger decay are shown in Fig.~\ref{fig:Kr_levels}.
The atom is initially in the ground state with energy~$E_0$.
Then, it absorbs an \XUV{}~photon with an energy of~$\omega\I{X}$.
This leads to the formation of a singly inner-shell ionized
cation with energy~$E^+$ and the ejection of a photoelectron.
With the Einstein relation, the peak of the energy distribution of the
photoelectron spectrum---the nominal photoelectron energy~$\Omega\I{P}$---is
found to be~$\vec k\I{P}^2 / 2 = \Omega\I{P}
= E_0 + \omega\I{X} - E^+$~\cite{Merzbacher:QM-98,Als-Nielsen:EM-01}.
The inner-shell hole Auger decays;
it is filled by a valence electron and the excess energy is transferred
ultrafast by electron correlations to a second valence electron which is
expelled.
This gives rise to an Auger line in the electron spectrum at the nominal
energy~$\vec k\I{A}^2 / 2 = \Omega\I{A} = E^+ - E^{++}$.
Afterwards, the system is in a dicationic final state with
energy~$E^{++}$.
This approximate mechanism of an ionization step with a following electronic
decay step is frequently referred to as two-step model
of Auger decay~\cite{Aberg:EP-82,Aberg:UT-92}.
Usually, this is a good approximation for atoms and molecules.
However, in condensed matter, rearrangement processes take place which
necessitate treating Auger decay as a one-step
process~\cite{Aberg:EP-82,Aberg:UT-92}.

Next, we consider the influence of an optical dressing laser.
Its intensity is assumed to be sufficiently low that it does not
excite or ionize atomic ground-state electrons.
However, the presence of the laser has a strong impact on the
outgoing electrons and this has important consequences for the observed signal.
We use Volkov waves~\cite{Wolkow:UK-35,Madsen:SF-05} to describe
continuum electrons and we exclude the possibility that the
initial \XUV{}~absorption does not ionize the atom but induces
only an excitation to a Rydberg orbital.
This means that effects such as electromagnetically induced transparency for
x~rays~\cite{Buth:TX-07,Buth:ET-07,Santra:SF-07,Buth:AR-08,%
Buth:RA-09,Buth:US-09} are not represented in our formalism.
Always, the laser intensity is assumed to be sufficiently low as not to modify
the Auger decay rate noticeably with respect to the laser-free case.
In other words, we neglect the impact of the laser on the
energies~$E_0$, $E^+$, and $E^{++}$ and the Auger decay-relevant
two-electron matrix elements (three atomic ground-state orbitals and
a continuum electron) because they are only weakly perturbed.

\subsection{Atomic electronic structure}
\label{sec:elstruct}

We assume a $Z$-electron atom with a closed-shell spin-singlet ground state.
The ground-state wave function is approximated by a Slater
determinant of one-electron orbitals~$\Phi_0(\vec r_1 \, \sigma_1,
\ldots, \vec r_Z \, \sigma_Z)$.
Electron coordinates are given by~$\vec r_i$ and spin projection
quantum numbers are given by~$\sigma_i$ for~$i \in \{1, \ldots, Z\}$.
We use the formalism of second quantization
where~$\hat b^{\vphantom{\dagger}}_{p\sigma}$ and
$\hat b^{\dagger}_{p\sigma}$ are
an annihilator and a creator of an electron in the spin
orbital~$\psi_{p\sigma}(\vec r)$, respectively, which is the tensor product
of a spatial orbital~$\varphi_p(\vec r)$ and a spinor of projection
quantum number~$\sigma$~\cite{Szabo:MQC-89}.
We define spin up $\sigma = \frac{1}{2} = {\uparrow}$ and spin down
$\sigma =-\frac{1}{2} = {\downarrow}$ and make the following ansatz for the
atomic ground state,
\begin{equation}
  \label{eq:SlaterDet}
  \ket{\Phi_0} = \prod^{Z/2}_{i=1} \hat
    b^{\dagger}_{i\uparrow} \, \hat b^{\dagger}_{i\downarrow} \ket{0} \; ,
\end{equation}
using the vacuum state~$\ket{0}$~\cite{Szabo:MQC-89}.

Let the full atomic electronic structure Hamiltonian be denoted
by~$\hat H\I{AT}$.
It consists of the kinetic energy of the electrons, the electron-nucleus
attraction, and the electron-electron repulsion~\cite{Szabo:MQC-89}.
We determine atomic orbitals within the
HFS~approximation~\cite{Slater:AS-51,Slater:XA-72} to~$\hat H\I{AT}$.
In a next step, we use the HFS~orbitals to represent~$\hat H\I{AT}$.
As the HFS~approximation is typically good, $\hat H\I{AT}$ naturally
decomposes into a part which is large and another which can be
treated as a perturbation.
We can rewrite~$\hat H\I{AT}$ as follows:
\begin{equation}
  \label{eq:atelst}
  \hat H\I{AT} = \hat H\I{HFS} + \hat H\I{CH} + \hat H\I{ee} \; ,
\end{equation}
assuming a representation in terms of the atomic orbitals.
We discuss these terms in the following paragraphs~\cite{Szabo:MQC-89}.

The HFS~Hamiltonian in Eq.~(\ref{eq:atelst}) represented in terms of
HFS~orbitals reads
\begin{equation}
  \label{eq:H_HFS}
  \hat H\I{HFS} = \Sum_{p} \varepsilon_p \> [
    \hat b^{\dagger}_{p \uparrow}   \, \hat b^{\vphantom{\dagger}}_{p \uparrow} +
    \hat b^{\dagger}_{p \downarrow} \, \hat b^{\vphantom{\dagger}}_{p \downarrow} ] \; ,
\end{equation}
with the spin-independent orbital energies~$\varepsilon_p$.
For notational clarity, we assume a countably infinite set of final states
in Eq.~(\ref{eq:H_HFS}) and in this entire section, \ie, we assume a finite
volume for box normalization of continuum wave
functions~\cite{Merzbacher:QM-98} or a (finite-element)
basis set expansion of the radial part of the
atomic orbitals times spherical harmonics~\cite{Buth:TX-07}.
From Sec.~\ref{sec:laser} onward, we will use continuum wave functions,
namely, plane waves and Volkov
waves~\cite{Wolkow:UK-35,Merzbacher:QM-98,Madsen:SF-05}.

The contribution~$\hat H\I{CH}$ in Eq.~(\ref{eq:atelst})
allows for the fact that in the Hartree-Fock-Slater
approximation~(\ref{eq:H_HFS}) the
electron-nuclear interaction and the electron-electron interaction
are replaced by the one-electron central potential~$V\I{HFS}(r)$ where the
nucleus is at the origin of a spherical polar coordinate
system and the electron is at radius~$r$~\cite{Buth:TX-07}.
This replacement is reversed by the corrective term
\begin{equation}
  \label{eq:H_CH}
  \hat H\I{CH} = \Sum_{p, q} \bra{\varphi_p} \hat h\I{CH} \ket{\varphi_q} \>
    [ \hat b^{\dagger}_{p \uparrow}   \, \hat b^{\vphantom{\dagger}}_{q \uparrow} +
      \hat b^{\dagger}_{p \downarrow} \, \hat b^{\vphantom{\dagger}}_{q \downarrow} ] \; .
\end{equation}
with
\begin{equation}
  \label{eq:one_h_CH}
  \hat h\I{CH} = -V\I{HFS}(r) - \frac{Z}{r}
\end{equation}
in addition to the explicit treatment of electron correlations
in the next paragraph.
There are no mixed terms involving a spin-up and a spin-down orbital in
Eq.~(\ref{eq:H_CH}) because the interaction~(\ref{eq:one_h_CH})
does not depend on the spin.

Electron correlations in Eq.~(\ref{eq:atelst}) are described by
\begin{equation}
  \label{eq:Hee}
  \hat H\I{ee} = \frac{1}{2} \Sum_{\sigma, \sigma', \xi, \xi' = \downarrow}
    ^{\uparrow} \Sum_{p, p', q, q'} V_{p\sigma \, p'\sigma' \, q\xi \,
    q'\xi'} \; \hat b^{          \dagger }_{p  \sigma } \,
               \hat b^{          \dagger }_{p' \sigma'} \,
               \hat b^{\vphantom{\dagger}}_{q' \xi'   } \,
               \hat b^{\vphantom{\dagger}}_{q  \xi    } \; .
\end{equation}
which is the only two-particle operator~\cite{Szabo:MQC-89,Merzbacher:QM-98}
in the total Hamiltonian of the atom.
The two-electron integrals in Eq.~(\ref{eq:Hee}) are defined in terms
of HFS~spin orbitals by~$V_{p\sigma \> p'\sigma' \, q\xi \, q'\xi'}
= \bra{\psi_{p\sigma} \, \psi_{p'\sigma'}} \hat h\I{ee} \ket{\psi_{q\xi}
\> \psi_{q'\xi'}}$.
The Coulomb interaction is
\begin{equation}
  \label{eq:twoelCoulomb}
  \hat h\I{ee} = \frac{1}{|\vec r - \vec r'|} \; ,
\end{equation}
where $\vec r$ and $\vec r'$ represent the coordinates of two
electrons in the atom.

\subsection{Atom in \XUV~light}
\label{sec:atomXUV}

We assume that the \XUV~light is linearly polarized along the
direction~$\vec e\I{X}$ and the wavelength is sufficiently large for
the electric dipole approximation to be adequate~\cite{Scully:QO-97}.
Then, the Hamiltonian for the interaction of electrons with
\XUV~light~\cite{Scully:QO-97} reads
\begin{equation}
  \label{eq:xuvint}
  \hat H\I{X} = \Sum_{p, q} D_{pq}(t) \,
    [ \hat b^{\dagger}_{p \uparrow}   \, \hat b^{\vphantom{\dagger}}_{q
      \uparrow} +
      \hat b^{\dagger}_{p \downarrow} \, \hat b^{\vphantom{\dagger}}_{q
      \downarrow} ] \; ,
\end{equation}
using the matrix element~$D_{pq}(t) \equiv \bra{\varphi_p} \hat h\I{X}
\ket{\varphi_q}$ in terms of atomic orbitals of the spin-independent
one-electron interaction with \XUV~light in length form,
\begin{equation}
  \label{eq:one_hX}
  \hat h\I{X} = \vec r \mul \vec E\I{X}(t) \; .
\end{equation}
We use for the electric field of the \XUV~light
\begin{equation}
  \label{eq:XUVfield}
  \vec E\I{X}(t) = \varepsilon\I{X}(t) \, \vec e\I{X}
    \cos(\omega\I{X} \, t) \; ,
\end{equation}
with the pulse envelope~$\varepsilon\I{X}(t)$ and the angular
frequency~$\omega\I{X}$.
With Eqs.~(\ref{eq:one_hX}) and (\ref{eq:XUVfield}), we can
rewrite the light-electron interaction as follows:
\begin{equation}
  \label{eq:XUVdipole}
  D_{pq}(t) = d_{pq} \, \frac{\varepsilon\I{X}(t)}{2} \, [ \euler^{\imag \,
    \omega\I{X} \, t} + \euler^{-\imag \, \omega\I{X} \, t} ] \; ,
\end{equation}
where the atomic dipole matrix elements are given by~$d_{pq} \equiv
\bra{\varphi_p} \vec r \mul \vec e\I{X} \ket{\varphi_q}$.

The complete Hamiltonian consists of the atomic electronic
structure~(\ref{eq:atelst}) and the interaction with the
\XUV{}~light~(\ref{eq:xuvint}) and reads
\begin{equation}
  \label{eq:hamcomplete}
  \hat H = \hat H\I{AT} + \hat H\I{X} \; .
\end{equation}
It will serve as the basis to treat the quantum dynamics of
\XUV{}~absorption and subsequent Auger decay.

\subsection{Excited states}
\label{sec:excitedstates}

Having formulated the Hamiltonian of the problem~(\ref{eq:hamcomplete})
and having specified the ground state~(\ref{eq:SlaterDet}), we need
to incorporate singly- and doubly- excited states in our description to
represent photoabsorption and Auger decay.
We use a single configuration-state function~(CSF), which is a linear
combination of singly- and doubly-excited determinants, to stand for
a singly- and a doubly-excited state, respectively.
Generally, singly-excited determinants are a satisfactory approximation to
singly-excited wave functions which is the reason for the success of
Koopmans' theorem~\cite{Szabo:MQC-89}.
However, doubly-excited states are not so well represented by
doubly-excited determinants
because of hole-hole repulsion effects and continuum electron interaction.
To overcome this approximation, one needs to allow for configuration
interaction~\cite{Szabo:MQC-89}.

Spin-singlet singly-excited states are represented
by~\cite{Szabo:MQC-89,Merzbacher:QM-98}
\begin{equation}
  \label{eq:singly}
  \ket{^1\Phi^a_h} = \frac{1}{\sqrt{2}} \> [\hat b^{\dagger}_{a
    \uparrow} \> \hat b^{\vphantom{\dagger}}_{h\uparrow}
     + \hat b^{\dagger}_{a\downarrow} \> \hat b^{\vphantom
    {\dagger}}_{h\downarrow}] \ket{\Phi_0}
  = \frac{1}{\sqrt{2}} \> [ \ket{\Phi^a_h} + \ket{\Phi^{\bar a}_{\bar h}}] \; .
\end{equation}
Here, $\hat b^{\vphantom{\dagger}}_{h \sigma}$~creates a hole in the
orbital~$h$ with spin projection number~$\sigma$ by destroying the
electron and $\hat b^{\dagger}_{a \sigma}$~creates an electron in the
orbital~$a$ with spin projection number~$\sigma$.
The hole orbital indices~$h$ which are taken into account form the
set~$\Cal H$;
the indices for virtual (unoccupied) orbitals are
$a > \frac{Z}{2}$ ($Z$~is even because we consider only closed-shell atoms).
After the second equals sign in Eq.~(\ref{eq:singly}), we introduce a
concise determinantial notation~\cite{Szabo:MQC-89}.
The bar over spatial orbital indices indicates a spin orbital with
spin down;
no bar refers to a spin orbital with spin up.

There are five classes of doubly-excited spin-singlet configuration-state
functions (Table~2.7 in Ref.~\onlinecite{Szabo:MQC-89}).
We focus on the two classes in which all
four spatial orbitals are distinct.
Using the concise notation of Eq.~(\ref{eq:singly}), we have
\begin{subeqnarray}
  \label{eq:doubly}
  \ket{\I{A}^1\Phi^{ab}_{ij}} &=& \frac{1}{\sqrt{12}} \> [
    2 \ket{\Phi^{ab}_{ij}} + 2 \ket{\Phi^{\bar a\bar b}_{\bar i\bar j}}
    - \ket{\Phi^{\bar ba}_{\bar ij}} \nonumber \\
  &&{} + \ket{\Phi^{\bar ab}_{\bar ij}}
    + \ket{\Phi^{a\bar b}_{i\bar j}} - \ket{\Phi^{b\bar a}_{i\bar j}} ] \; , \\
  \ket{\I{B}^1\Phi^{ab}_{ij}} &=& \frac{1}{2} \> [
    \ket{\Phi^{\bar ba}_{\bar ij}} + \ket{\Phi^{\bar ab}_{\bar ij}}
    + \ket{\Phi^{a\bar b}_{i\bar j}} + \ket{\Phi^{b\bar a}_{i\bar j}} ] \; .
\end{subeqnarray}
The pairs of orbital indices~$(i,j)$ of double vacancies with~$i < j$,
which are considered, constitute the set~$\Cal F$.
With the restrictions~$a < b$ and $a,b > \frac{Z}{2}$ for the
virtual orbital indices, we enumerate all distinct doubly-excited
configurations.

The energy of the ground state~(\ref{eq:SlaterDet}) is found with
the electronic Hamiltonian~(\ref{eq:atelst}) as follows:
\begin{equation}
  \label{eq:groundenergy}
  E_0 = \bra{\Phi_0} \hat H\I{AT} \ket{\Phi_0} = 2 \Sum_{i=1}^{Z/2}
    \varepsilon_i + \bra{\Phi_0} \hat H\I{CH} + \hat H\I{ee} \ket{\Phi_0} \; .
\end{equation}
The first term on the right-hand side is twice the sum of occupied
Hartree-Fock-Slater orbital energies dubbed~$\Cal E_0$~\cite{Szabo:MQC-89}.
The singly-excited states~(\ref{eq:singly}) have an energy of
\begin{equation}
  \label{eq:singlyenergy}
  E^a_h = \bra{^1\Phi^a_h} \hat H\I{AT} \ket{^1\Phi^a_h} = \Cal E_0
    - \varepsilon_h + \varepsilon_a + \bra{^1\Phi^a_h} \hat H\I{CH}
    + \hat H\I{ee} \ket{^1\Phi^a_h} \; ,
\end{equation}
and the doubly-excited states~(\ref{eq:doubly}) have an energy of
\begin{equation}
  \label{eq:doublyenergy}
  \begin{array}{rcl}
  \displaystyle E^{ab}_{ij} &\equiv& \displaystyle E^{ab}_{ij,\mathrm{A}}
    = E^{ab}_{ij,\mathrm{B}} = \bra{\I{A}^1\Phi^{ab}_{ij}} \hat H\I{AT}
    \ket{\I{A}^1\Phi^{ab}_{ij}} \\
  &=& \displaystyle \Cal E_0 - \varepsilon_i - \varepsilon_j
    + \varepsilon_a + \varepsilon_b + \bra{\I{A}^1\Phi^{ab}_{ij}}
    \hat H\I{CH} + \hat H\I{ee} \ket{\I{A}^1\Phi^{ab}_{ij}} \; .
  \end{array}
\end{equation}
We have~$E^{ab}_{ij,\mathrm{A}} = E^{ab}_{ij,\mathrm{B}}$ because both
doubly-excited CSFs consist of determinants with excitations from the
same spatial orbitals into the same spatial orbitals, only the
spinors change.
As our nonrelativistic Hamiltonian~(\ref{eq:atelst}) does not
depend on spin, especially, it does not contain spin-orbit coupling,
the energies are the same for all the excited determinants and
thus also for the configuration-state functions~A and~B.

\subsection{Simplified Hamiltonian}
\label{sec:modelHam}

We have formulated a full \emph{ab initio} description of the problem
with a truncated excitation manifold~\cite{Szabo:MQC-89}.
Our framework represents an ideal starting point for further simplifications.
Eventually, it will be reduced to an essential-states model in
Sec.~\ref{sec:essstmo} that contains only the absolutely
necessary energies and matrix elements to still describe the physics
of the processes.
We form a matrix representation of the Hamiltonian~$\hat H$
[Eq.~(\ref{eq:hamcomplete})] in terms of the orthonormal basis,
\begin{equation}
  \label{eq:basis}
  \begin{array}{rcl}
    \displaystyle \Cal B &=&\displaystyle \{ \ket{\Phi_0}, \ket{^1\Phi^a_h},
      \ket{_{\gamma}^1 \Phi^{ab}_{ij}} \  {|} \  h \in \Cal H,
      (i,j) \in \Cal F, \\
    &&\displaystyle \qquad a, b > Z/2, a < b, \gamma \in \{\mathrm{A,B}\}
      \} \; .
  \end{array}
\end{equation}
We decompose the representation of~$\hat H$
into an exactly solvable part~$\hat H_0$, given by the diagonal elements
of~$\hat H$, and a perturbation~$\hat H_1$,
given by the off-diagonal elements of~$\hat H$.
This is a so-called Epstein-Nesbet
partitioning~\cite{Epstein:SE-26,Nesbet:CI-55,Buth:NH-04}.
The exactly solvable diagonal part is written compactly in
first quantization as
\begin{equation}
  \label{eq:H_0}
  \begin{array}{rcl}
  \displaystyle \hat H_0 &=& \displaystyle \ket{\Phi_0} E_0 \bra{\Phi_0} +
    \sum_{h \in \Cal H} \Sum_{a > Z/2} \ket{^1\Phi^a_h} E^a_h
    \bra{^1\Phi^a_h} \\
    &&\displaystyle{}+ \sum_{(i,j) \in \Cal F} \sum_{\atopa{a,b > Z/2}{a < b}}
    \sum_{\gamma \in \{\mathrm{A,B}\}} \ket{_{\gamma}^1\Phi^{ab}_{ij}}
    E^{ab}_{ij,\gamma} \bra{_{\gamma}^1\Phi^{ab}_{ij}} \; ,
  \end{array}
\end{equation}
where we use the energies from Eqs.~(\ref{eq:groundenergy}),
(\ref{eq:singlyenergy}), and (\ref{eq:doublyenergy}).
These energies are well suited to be treated as (experimental)
parameters [see Sec.~\ref{sec:compdet}].
Here, $\hat H\I{X}$ makes no contribution to the diagonal matrix elements
of~$\hat H$ because it consists only of off-diagonal elements.

The perturbation is given by~$\hat H_1$.
We do not use all off-diagonal elements of~$\hat H$
for~$\hat H_1$ and, additionally, we make approximations to
the ones we retain.
We use
\begin{equation}
  \label{eq:H_1}
  \begin{array}{rcl}
    \hat H_1 &=& \Sum_{h \in \Cal H} \Sum_{a > Z/2} \bigl[
      \ket{\Phi_0} \bra{\Phi_0} \hat H\I{X} \ket{^1\Phi^a_h} \bra{^1\Phi^a_h} \\
    &&{} + \ket{^1\Phi^a_h} \bra{^1\Phi^a_h} \hat H\I{X} \ket{\Phi_0}
      \bra{\Phi_0} \bigr] \\
    &&{} + \Sum_{\atopa{h \in \Cal H}{(i,j) \in \Cal F}} \Sum_{\atopa{a,b > Z/2}
      {a < b}} \Sum_{\gamma \in \{\mathrm{A,B}\}} \bigl[
      \ket{^1\Phi^a_h} \bra{^1\Phi^a_h} \hat H\I{ee}
      \ket{_{\gamma}^1\Phi^{ab}_{ij}} \bra{_{\gamma}^1\Phi^{ab}_{ij}} \\
    &&{} + \ket{_{\gamma}^1\Phi^{ab}_{ij}} \bra{_{\gamma}^1\Phi^{ab}_{ij}}
      \hat H\I{ee} \ket{^1\Phi^a_h} \bra{^1\Phi^a_h} \bigr] \; .
  \end{array}
\end{equation}
Here, $\hat H\I{HFS}$ [Eq.~(\ref{eq:H_HFS})] makes no contribution
because it consists only of diagonal elements.
The impact of $\hat H\I{CH}$ [Eq.~(\ref{eq:H_CH})] is neglected totally
and we consider the energies in Eq.~(\ref{eq:H_0}) to be parameters
which shall compensate for this and the other omissions that influence
the energies of the involved states.
Similarly, electron correlations~$\hat H\I{ee}$ are only included when
they cause transitions between singly- and doubly-excited states,
\ie, they are taken into account when they are responsible
for Auger decay which cannot be understood in a mean-field picture.
In principle, the neglected matrix elements can be incorporated
allowing one to carry out a full \emph{ab initio} treatment of the problem.

\subsection{Equation of motion formulation of photoionization and Auger decay}
\label{sec:EOM}

To describe a photoionization process with subsequent Auger decay,
we solve the time-dependent Schr\"odinger equation
\begin{equation}
  \label{eq:tdschroedi}
  \hat H \ket{\Psi,t} = \imag \frac{\partial}{\partial t} \ket{\Psi,t}
\end{equation}
for an atom in \XUV~light.
In terms of the states in the basis~$\Cal B$ [Eq.~(\ref{eq:basis})],
a general state ket (or wave packet) is given by
\begin{equation}
  \label{eq:wavepacket}
  \begin{array}{rcl}
  \displaystyle \ket{\Psi,t} &=& c_0(t) \> \euler^{-\imag E_0 t} \ket{\Phi_0}
    + \Sum_{\atopa{h \in \Cal H}{a > Z/2}} c^a_h(t) \> \euler^{-\imag E^a_h t}
    \ket{^1\Phi^a_h} \\
  &&\displaystyle{}+ \Sum_{\atopa{(i,j) \in \Cal F}{\gamma \in
    \{\mathrm{A,B}\}}} \Sum_{\atopa{a,b > Z/2}{a < b}} c^{ab}_{ij,\gamma}(t) \>
    \euler^{-\imag E^{ab}_{ij,\gamma} t} \ket{_{\gamma}^1\Phi^{ab}_{ij}} \; ,
  \end{array}
\end{equation}
which we insert into the time-dependent Schr\"odinger
equation~(\ref{eq:tdschroedi}).
Exploiting~$\hat H_0 \ket{\phi} = E_{\phi} \ket{\phi}$ for states~$\ket{\phi}
\in \Cal B$ with energies~$E_{\phi}$, we arrive at the EOMs for the expansion
coefficients~$c_0(t)$, $c^a_h(t)$, and $c^{ab}_{ij,\gamma}(t)$
by projecting on~$\bra{\phi}$ for all~$\ket{\phi} \in \Cal B$.
The atom is initially in the ground state which implies
the initial conditions~$c_0(0) = 1$ and $c^a_h(0) = c^{ab}_{ij,\gamma}(0) = 0$.

We get the first EOM for~$\bra{\phi} = \bra{\Phi_0}$ which represents
the rate of change of the ground-state amplitude,
\begin{equation}
  \label{eq:EOMground}
  \dot c_0(t) = -\imag \, \sqrt{2} \, \Sum_{h \in \Cal H} \Sum_{a > Z/2}
    D_{ha}(t) \, \euler^{\imag \, (E_0 - E^a_h) \, t} \, c^a_h(t) \; .
\end{equation}
We consider here the weak absorption limit, \ie, $c_0(t) \approx 1$ for
all times.
The rate of change will, nevertheless, prove highly beneficial in
determining the cross section in Sec.~\ref{sec:xsect} and the photoelectron
spectrum in Sec.~\ref{sec:ldpel}.

The second EOM results from~$\ket{\phi} = \ket{^1\Phi^a_h}$ and
describes the inner-shell hole amplitude for~$h \in \Cal H$,
\begin{equation}
  \label{eq:EOMxuv}
  \begin{array}{rcl}
  \displaystyle \dot c^a_h(t) &=& \displaystyle -\imag \, \sqrt{2} \,
    D_{ah}(t) \, \euler^{\imag (E^a_h - E_0) t} \\
  &&\displaystyle{}- \imag \, \Sum_{(i,j) \in \Cal F}
    \Sum_{\atopa{b > Z/2}{a \neq b}} \Bigl [ -\sqrt{\frac{3}{2}} \,
    v^*_{hb[ij]} \> \euler^{\imag (E^a_h
    - E^{ab}_{ij,\mathrm{A}}) t} \, c^{ab}_{ij,\mathrm{A}}(t) \\
  &&\displaystyle{}-{\frac{1}{\sqrt{2}}} \, v^*_{hb\{ij\}} \>
    \euler^{\imag (E^a_h - E^{ab}_{ij,\mathrm{B}}) t} \,
    c^{ab}_{ij,\mathrm{B}}(t) \Bigr ] \; .
  \end{array}
\end{equation}
The first term on the right-hand side of the equation represents hole
production due to absorption of \XUV{}~light;
the second term describes the loss of hole amplitude caused by
Auger decay.
In this equation, the two-electron matrix element in terms of spatial
(\eg, Hartree-Fock-Slater) atomic orbitals is denoted by
\begin{equation}
  \label{eq:twospatial}
  v_{pp'qq'} = \bra{\varphi_p \> \varphi_{p'}} \hat h\I{ee}
    \ket{\varphi_q \> \varphi_{q'}} \; .
\end{equation}
Further, we define the antisymmetrized two-electron matrix
element~$v_{pp'[qq']} = v_{pp'qq'} - v_{pp'q'q}$ and the symmetrized
two-electron matrix element~$v_{pp'\{qq'\}} = v_{pp'qq'}
+ v_{pp'q'q}$ which consist of a direct matrix element~$v_{pp'qq'}$ and
an exchange matrix element~$v_{pp'q'q}$.

The third EOM is obtained setting~$\bra{\phi} = \bra{_{\gamma}^1
\Phi^{ab}_{ij}}$;
it describes the Auger decay amplitude
\begin{subeqnarray}
  \label{eq:EOMauger}
  \slabel{eq:EOMaugerA}
  \dot c^{ab}_{ij,\mathrm{A}}(t) &=& \imag \, \sqrt{\frac{3}{2}} \
    \Sum_{h \in \Cal H} v_{hb[ij]} \> \euler^{\imag
    (E^{ab}_{ij,\mathrm{A}} - E^a_h) t} \, c^a_h(t) \; , \\
  \slabel{eq:EOMaugerB}
  \dot c^{ab}_{ij,\mathrm{B}}(t) &=& \imag \, {\frac{1}{\sqrt{2}}} \
    \Sum_{h \in \Cal H} v_{hb\{ij\}} \> \euler^{\imag
    (E^{ab}_{ij,\mathrm{B}} - E^a_h) t} \, c^a_h(t) \; .
\end{subeqnarray}
We can reduce the EOMs~(\ref{eq:EOMxuv}) and (\ref{eq:EOMauger}) further
by ignoring the electron exchange matrix element~$v_{hbji}$ and exploiting
the fact that the energies of the doubly-excited states~(\ref{eq:doublyenergy})
are the same, $E^{ab}_{ij} \equiv E^{ab}_{ij,\mathrm{A}}
= E^{ab}_{ij,\mathrm{B}}$.
Then, the second EOM~(\ref{eq:EOMxuv}) for the inner-shell hole amplitude
simplifies to
\begin{equation}
  \label{eq:EOMxuvOlga}
  \begin{array}{rcl}
  \displaystyle \dot c^a_h(t) &=& \displaystyle -\imag \, \sqrt{2}
    \, D_{ah}(t) \, \euler^{\imag (E^a_h - E_0) t} \\
  &&\displaystyle{}+ \imag \, 2 \, \sqrt{2} \, \Sum_{(i,j) \in \Cal F}
    \Sum_{\atopa{b > Z/2}{a \neq b}} v^*_{hbij} \>
    \euler^{\imag (E^a_h - E^{ab}_{ij}) t} \, c^{ab}_{ij}(t) \; ,
  \end{array}
\end{equation}
for~$h \in \Cal H$ with the definition~$c^{ab}_{ij}(t) \equiv
c^{ab}_{ij,\mathrm{B}}(t)$.
Again ignoring electron exchange and using~$E^{ab}_{ij}$,
leads us to the relation~$c^{ab}_{ij,\mathrm{A}}(t) = \sqrt{3} \,
c^{ab}_{ij,\mathrm{B}}(t)$ between Eqs.~(\ref{eq:EOMaugerA}) and
(\ref{eq:EOMaugerB});
thereby, we use that both $c^{ab}_{ij,\mathrm{A}}(t)$
and $c^{ab}_{ij,\mathrm{B}}(t)$ vanish initially.
Then, we need to retain only the simplified Eq.~(\ref{eq:EOMaugerB}) of
the two EOMs for the Auger decay amplitude~(\ref{eq:EOMauger}) yielding
\begin{equation}
  \label{eq:EOMaugerOlga}
  \dot c^{ab}_{ij}(t) = \imag \, {\frac{1}{\sqrt{2}}} \
    \Sum_{h \in \Cal H} v_{hbij} \> \euler^{\imag
    (E^{ab}_{ij} - E^a_h) t} \, c^a_h(t) \; .
\end{equation}
Equations~(\ref{eq:EOMxuvOlga}) and (\ref{eq:EOMaugerOlga}) constitute a
linear system of differential equations which contains all phase
information and thus describes interference effects.

\section{Laser dressing}
\label{sec:laserdre}

In Sec.~\ref{sec:auger}, we devised a formalism to describe the
quantum dynamics of the photoionization of the inner shell of an atom
by \XUV{}~light and the subsequent Auger decay.
Here we expand our formalism to include an additional optical dressing
laser of moderate intensity.
The impact of the laser on ground-state electrons is neglected and
only the modification of the continuum wave functions of the
photo- and the Auger electron is considered.

To begin with, we simplify the manifold of virtual states
by replacing it by free-electron wave functions,
the momentum normalized plane waves~\cite{Merzbacher:QM-98},
\begin{equation}
  \label{eq:planewave}
  \varphi_{\mathrm P, \vec k}(\vec r, t) = \frac{1}{(2\pi)^{3/2}} \>
    \euler^{\imag \, \bigl( \vec k \mul \vec r -
    \frac{\vec k^2}{2} \, t \bigr)} \; .
\end{equation}
This substitution explicitly excludes Rydberg states.
It is justified by the fact that we are only concerned with continuum
electrons of sizable kinetic energy.
A consequence of our replacement is that the new continuum wave functions
are no longer strictly orthogonal to the bound-state wave functions
because both sorts of wave functions stem from different
Hamiltonians:~$\frac{\hat{\vec p}^2}{2} = -\frac{\vec \nabla^2}{2}$
and $\hat H\I{HFS}$ [Eq.~(\ref{eq:H_HFS})]~\cite{Merzbacher:QM-98}.
Note that we exploited strict orthogonality in the derivations of
Sec.~\ref{sec:auger}.

When we additionally consider a laser field, our replacement of
continuum wave functions becomes known as strong-field
approximation~\cite{Madsen:SF-05} in which the influence of the Coulomb
potential on continuum states is neglected.
In other words, laser dressing can be incorporated easily into our
treatment by replacing the
laser-free continuum functions~$\varphi_{\mathrm P, \vec k}(\vec r, t)$
by Volkov waves~\cite{Wolkow:UK-35,Madsen:SF-05} [see
Eq.~(\ref{eq:VolkovState}) below].
In doing so, electrons in the atomic ground-state orbitals, however,
are considered to be uninfluenced by the laser.

The laser pulse is assumed to be long with respect to all other time
scales in this paper and is taken to be monochromatic and
continuous wave.
Let the laser radiation of angular frequency~$\omega\I{L}$ be linearly
polarized with the polarization vector~$\vec e\I{L}$.
The vector potential is
\begin{equation}
  \label{eq:lasVecPot}
  \vec A\I{L}(t) = \vec A\I{L,0} \, \sin(\omega\I{L} \, (t + \Delta t)) \; ,
\end{equation}
where the amplitude is~$\vec A\I{L,0} = -{\Cal A}\I{L} \, \vec e\I{L}$.
The laser phase at~$t = 0$ can be specified using~$\Delta t$.
Then, the laser electric field follows from
\begin{equation}
  \label{eq:lasElField}
  \vec E\I{L}(t) = -\frac{\differential \vec A\I{L}(t)}{\differential t}
                 = \vec E\I{L,0} \, \cos(\omega\I{L} \, (t + \Delta t)) \; ,
\end{equation}
where the electric-field amplitude is~$\vec E\I{L,0} = \omega\I{L} \,
\Cal A\I{L} \, \vec e\I{L}$ [see Eq.~(\ref{eq:XUVfield})].
We assume free fields, \ie, a vanishing scalar potential,
and the Coulomb gauge~\cite{Scully:QO-97,Madsen:SF-05}.

The Hamiltonian of a free electron in a laser field in velocity
form~\cite{Scully:QO-97,Madsen:SF-05} is
\begin{equation}
  \label{eq:FreeLaserHam}
  \hat h\I{V} = \dfrac{(\hat{\vec p} + \vec A\I{L}(t))^2}{2} \; ,
\end{equation}
using the electron momentum operator~$\hat{\vec p} = - \imag \vec \nabla$.
The solution of the time-dependent Schr\"odinger equation~(\ref{eq:tdschroedi})
with Hamiltonian~(\ref{eq:FreeLaserHam}) reads
\begin{equation}
  \label{eq:VolkovState}
  \varphi_{\mathrm V, \vec k}(\vec r, t) = \frac{1}{(2\pi)^{3/2}} \>
    \euler^{\imag \, (\vec k \mul \vec r - \Phi\I{V}(\vec k, t))}
\end{equation}
and is called a Volkov
wave~\nocite{Scully:QO-97}\cite{Wolkow:UK-35,Madsen:SF-05,footnote1}.
The Volkov phase is given by
\begin{equation}
  \label{eq:VolkovPhase}
  \Phi\I{V}(\vec k, t) = \frac{1}{2} \Int^{t}_{-\infty} [\vec k +
    \vec A\I{L}(t')]^2 \, \euler^{\eta \, t'} \differential t' \; .
\end{equation}
The factor~$\euler^{\eta \, t'}$ for~$\eta > 0$ ensures
the convergence of the integral.
After the integral has been performed, the limit~$\eta \to 0^+$ is taken.
The Volkov phase vanishes at~$-\infty$.
Inserting Eq.~(\ref{eq:lasVecPot}) into Eq.~(\ref{eq:VolkovPhase}),
we find
\begin{equation}
  \label{eq:VolkovPhaseCW}
  \begin{array}{rcl}
  \displaystyle \Phi\I{V}(\vec k, t) &=& \displaystyle \Bigl( \frac{\vec
    k^2}{2} + U\I{P} \Bigr) \; t - \vec k \mul \vec \alpha\I{L} \>
    \cos(\omega\I{L} \, (t + \Delta t)) \\
  &&\displaystyle{} - \frac{U\I{P}}{2 \, \omega\I{L}} \> \sin(2 \, \omega\I{L}
    \, (t + \Delta t)) \; ,
  \end{array}
\end{equation}
where the ponderomotive potential is
\begin{equation}
  \label{eq:pondpot}
  U\I{P} = \frac{\Cal A\I{L}^2}{4} = \frac{2\pi}{\omega^2\I{L}} \>
  \alpha \, I\I{L,0} \; ,
\end{equation}
with the electric-field amplitude~$|\vec E\I{L,0}| = \sqrt{8 \pi \,
\alpha \, I\I{L,0}}$ for a laser with intensity~$I\I{L,0}$.
The fine-structure constant is~$\alpha$.
During a laser cycle, the maximum classical excursion from the origin of a
free electron is given by~\cite{Madsen:SF-05}
\begin{equation}
  \label{eq:maxexcurs}
  \vec \alpha\I{L} = \frac{\vec A\I{L,0}}{\omega\I{L}} = -\frac{\sqrt{8\pi
    \, \alpha \, I\I{0,L}}}{\omega\I{L}^2} \> \vec e\I{L} \; .
\end{equation}

The exponential of the Volkov phase~(\ref{eq:VolkovPhaseCW}),
$\euler^{-\imag \, \Phi\I{V}(\vec k, t)}$, can be expanded using the
generating function of the generalized Bessel
functions~$J_m(u,v)$~\cite{Reiss:EI-80,Madsen:SF-05} which reads
\begin{equation}
  \label{eq:genBesgen}
  \euler^{-\imag \, (u \, \cos \theta + v \, \sin (2 \, \theta))}
    = \Sum^{\infty}_{m=-\infty} (-\imag)^m \>
    \euler^{\imag \, m \, \theta} \> J_m(u,v) \; ,
\end{equation}
by setting~$\phi = \theta - \frac{\pi}{2}$ in Eq.~(10) in
Ref.~\onlinecite{Madsen:SF-05}.
The~$J_m(u,v)$ can be evaluated in terms of the ordinary Bessel
functions~$J_{m-2n}(u)$ and $J_n(v)$ using~\cite{Reiss:EI-80,Madsen:SF-05}
\begin{equation}
  \label{eq:genBesDef}
  J_m(u,v) = \Sum_{n=-\infty}^{\infty} J_{m-2n}(u) \, J_n(v) \; .
\end{equation}
Another connection to ordinary Bessel functions represents the
relation~$J_m(u) = J_m(u,0)$ with which Eq.~(\ref{eq:genBesgen})
reduces to the familiar Jacobi-Anger expansion~\cite{Arfken:MM-05}.
Finally, the temporal phase factor reduces to
\begin{equation}
  \label{eq:VolkovExpand}
  \begin{array}{rcl}
  \displaystyle \euler^{-\imag \, \Phi\I{V}(\vec k, t)} &=& \displaystyle
    \euler^{-\imag \, ( \frac{\vec k^2}{2} + U\I{P} ) \, t}
    \Sum^{\infty}_{m=-\infty} \euler^{-\imag \, m \, (\omega\I{L}
    \, (t + \Delta t) - \frac{\pi}{2})} \\
  &&\displaystyle\hspace{7em} \times J_m \bigl( \vec \alpha\I{L} \mul
    \vec k, \frac{U\I{P}}{2 \, \omega\I{L}} \bigr) \; .
  \end{array}
\end{equation}

\section{Essential-states model for laser-dressed photoionization and
Auger decay}
\label{sec:essstmo}

This section is devoted to a solution of the EOMs from Sec.~\ref{sec:EOM}
for an essential-states
model~\cite{Fedorov:AF-97,Smirnova:QC-03,Yakovlev:ES-03} which is a
fairly drastic approximation to the model Hamiltonian of
Sec.~\ref{sec:modelHam}.
However, it retains the essential physics of the problem.
Namely, we include only the states from three magnetic subshells of
the occupied orbital manifold and form essential states from them
by averaging the dipole and two-electron matrix elements in
Sec.~\ref{sec:maelen} over the subshells.
Subsequently, we decouple the system of EOMs in
Sec.~\ref{sec:pertdec} using perturbation theory.
The resulting equations are adapted to account for laser dressing following
Sec.~\ref{sec:laserdre} and are solved analytically.
We obtain the hole-state amplitude in Sec.~\ref{sec:hostamp}
which is used to find the Auger electron spectrum in Sec.~\ref{sec:laser}.
Finally, we derive the \XUV{}~absorption cross section in
Sec.~\ref{sec:xsect} and the photoelectron spectrum in Sec.~\ref{sec:ldpel}.

\subsection{Matrix elements and energies}
\label{sec:maelen}

To begin with, let us simplify the problem and the notation.
We represent photo- and Auger electrons by plane waves~(\ref{eq:planewave})
with momentum vectors~$\vec k\I{P}$ and $\vec k\I{A}$, respectively
(laser dressing is not treated at this point).
Moreover, we make the following replacements:
$a \to \vec k\I{P}$ and $b \to \vec k\I{A}$ in our previous equations.
Summations over~$a$ and $b$ become integrals over~$\vec k\I{P}$
and $\vec k\I{A}$.
As $h$, $i$, and $j$ refer to individual orbitals, their use is not
meaningful anymore in our model context and are eliminated or
adequately substituted as detailed in the following.

To construct the matrix elements of the essential-states model,
we note that the set of hole orbital indices~$\Cal H$
refers to orbitals
from a single magnetic subshell.
The pairs~$(i,j) \in \Cal F$ refer to orbital~$i$~from one magnetic subshell
and orbital~$j$ from another magnetic subshell.
The number of single hole states is~$\# \Cal H$ and the number of double
hole states is~$\# \Cal F$.
The \XUV{}~interaction matrix element is taken to be
\begin{equation}
  \label{eq:DipME}
  \bar d(\vec k\I{P}) = Q_d \, \sqrt{\frac{1}{\# \Cal H}
    \Sum_{h \in \Cal H} |d_{\vec k\I{P} \, h}|^2}
\end{equation}
while the Auger decay matrix element is
\begin{equation}
  \label{eq:AugME}
  \bar v(\vec k\I{A}) = Q_v \, \sqrt{\frac{1} {\# \Cal H \; \# \Cal F}
    \Sum_{h \in \Cal H} \Sum_{(i,j) \in \Cal F}
    |v_{h \, \vec k\I{A} \, i \, j}|^2} \; .
\end{equation}
Here, $Q_d > 0$~is the strength of the dipole matrix element
and $Q_v > 0$~is the strength of the two-electron matrix element.
Both strengths will be determined later in
Eqs.~(\ref{eq:strengthdipole}) and (\ref{eq:strengthtwoel}), respectively,
based on (experimental) parameters.
We chose to use the rms~value to form average matrix elements because
in the following equations frequently the modulus squared
of the matrix elements is used.

Within the scope of our essential-states model,
the energies of the states of Sec.~\ref{sec:excitedstates}
and Fig.~\ref{fig:Kr_levels} are as follows.
The ground-state energy~(\ref{eq:groundenergy}) is~$E_0 = \Cal E_0$,
neglecting the influence of~$\hat H\I{CH} + \hat H\I{ee}$.
The energy of singly-excited states~(\ref{eq:singlyenergy})
is decomposed into the kinetic energy of the
photoelectron~$\frac{\vec k\I{P}^2}{2}$ and the energy of the cation~$E^+$.
It becomes~$E^a_h \to \frac{\vec k\I{P}^2}{2}
+ E^+$ with~$E^+ = \Cal E_0 - \varepsilon_h$ (the orbital indices
in~$\Cal H$ denote orbitals from the same subshell and
thus~$\varepsilon_h$~is the same for all~$h \in \Cal H$).
For doubly-excited states~(\ref{eq:doublyenergy}), we
set~$E^{ab}_{ij} \to \frac{\vec k\I{P}^2 + \vec k\I{A}^2}{2}
+ E^{++}$ with $E^{++} = \Cal E_0 - \varepsilon_i
- \varepsilon_j$~being the energy of the dication
[again $\varepsilon_i$ and $\varepsilon_j$~are the same for
all~$(i,j) \in \Cal F$].
With these definitions, we find for the single ionization
potential~$I^+ = E^+ - E_0 = - \varepsilon_h$ and for the double ionization
potential~$I^{++} = E^{++} - E_0 = - \varepsilon_i - \varepsilon_j$
[see also Fig.~\ref{fig:Kr_levels}].
The nominal photoelectron energy from the stationary-state energy level
scheme is~$\Omega\I{P} = E_0 - E^+ + \omega\I{X}$.
Likewise, $\Omega\I{A} = E^+ - E^{++}$ is the nominal
Auger electron energy.

Under these assumptions, we find for the inner-shell hole
amplitude~(\ref{eq:EOMxuvOlga})---in which we replaced the hole index~$h$,
by the subscript~P and the double index~$ij$ by the subscript~A---the
expression
\begin{equation}
  \label{eq:EOMxuvOlgaFin}
  \begin{array}{rcl}
  \displaystyle \dot{\bar c}\I{P}^{\vec k\I{P}}(t) &=& \displaystyle
    \frac{-\imag}{\sqrt{2}} \; \bar d(\vec k\I{P}) \, \varepsilon\I{X}(t) \,
    \euler^{\imag (\frac{\vec k\I{P}^2}{2} - \Omega\I{P}) t} \\
  &&\displaystyle{} + \underbrace{\imag \, 2 \, \sqrt{2}
    \Int_{\mathbb{R}^3} \bar v^*(\vec k\I{A}) \> \euler^{-\imag
    (\frac{\vec k\I{A}^2}{2} - \Omega\I{A}) t} \, \bar c\I{A}^{\vec k\I{P}
    \, \vec k\I{A}}(t) \differential^3 k\I{A}}_{\textstyle
    \equiv w(\vec k\I{P}, t)} \; .
  \end{array}
\end{equation}
The Auger decay amplitude~(\ref{eq:EOMaugerOlga}) is
\begin{equation}
  \label{eq:EOMaugerOlgaFin}
  \dot{\bar c}\I{A}^{\vec k\I{P} \, \vec k\I{A}}(t) =
    \frac{\imag}{\sqrt{2}} \>
    \bar v(\vec k\I{A}) \>
    \euler^{\imag (\frac{\vec k\I{A}^2}{2} - \Omega\I{A}) t} \, \bar
    c\I{P}^{\vec k\I{P}}(t) \; .
\end{equation}
The EOMs of the essential-states model are formed by
Eqs.~(\ref{eq:EOMxuvOlgaFin}) and (\ref{eq:EOMaugerOlgaFin}).
They are very similar to Eqs.~(10) and (11) in
Ref.~\cite{Smirnova:QC-03,footnote2}.

\subsection{Perturbative decoupling}
\label{sec:pertdec}

Despite our considerable simplifications in Sec.~\ref{sec:maelen},
Eqs.~(\ref{eq:EOMxuvOlgaFin}) and (\ref{eq:EOMaugerOlgaFin}) still form
a linear system of coupled integrodifferential equations.
The coupling stems from the second term~$w(\vec k\I{P}, t)$
on the right-hand side of Eq.~(\ref{eq:EOMxuvOlgaFin}).
It describes the Auger decay of inner-shell holes
and can be approximated in terms of second-order time-dependent
perturbation theory (Weisskopf-Wigner
theory)~\cite{Weisskopf:-30,Sakurai:MQM-94,Merzbacher:QM-98}.
This treatment allows us to decouple the differential
equation~(\ref{eq:EOMxuvOlgaFin}) by eliminating the
dependence on~$\bar c\I{A}^{\vec k\I{P} \,
\vec k\I{A}}(t)$ in terms of a decay width~$\Gamma$ and an energy
shift~$\Delta_{\mathrm{R}}$ as follows:
\begin{equation}
  \label{eq:WWdecay}
  w(\vec k\I{P}, t) =
    \Bigl( -\imag \Delta_{\mathrm{R}} - \frac{\Gamma}{2} \Bigr) \; \bar
    c\I{P}^{\vec k\I{P}}(t) \; .
\end{equation}
The energy shift of the resonance state follows from
\begin{equation}
  \label{eq:resenshift}
  \begin{array}{rcl}
    \displaystyle \Delta_{\mathrm{R}, h} &=& \displaystyle \Sum_{(i,j) \in
      \Cal F} \Sum_{\gamma \in \{\mathrm{A,B}\}}
      \Pr \Int_{\mathbb{R}^3} \frac{|\bra{_{\gamma}^1
      \Phi^{\vec k\I{P} \, \vec k\I{A}}_{ij}} \hat H_1
      \ket{^1\Phi^{\vec k\I{P}}_h}|^2}{E^{\vec k\I{P}}_h -
      E^{\vec k\I{P} \, \vec k\I{A}}_{ij, \gamma}} \differential^3 k\I{A} \\
    &\approx& \displaystyle \Sum_{(i,j) \in \Cal F}
      \Pr \Int_{\mathbb{R}^3} \frac{2 \, |v_{h\,\vec k\I{A}\,i\,j}|^2}
      {E^{\vec k\I{P}}_h - E^{\vec k\I{P} \, \vec k\I{A}}_{ij}}
      \differential^3 k\I{A} \; ,
  \end{array}
\end{equation}
where $\Pr$~indicates that the principle value of the integral has to be
taken.
The result was obtained by neglecting electron exchange and
using~$E^{\vec k\I{P} \, \vec k\I{A}}_{ij} \equiv
E^{\vec k\I{P} \, \vec k\I{A}}_{ij,\mathrm A} =
E^{\vec k\I{P} \, \vec k\I{A}}_{ij,\mathrm B}$.
With the same assumptions, the decay width becomes
\begin{equation}
  \label{eq:WWdecwidth}
  \begin{array}{rcl}
    \displaystyle \Gamma_h &=& \displaystyle 2 \pi \Sum_{(i,j) \in \Cal F}
      \Sum_{\gamma \in \{\mathrm{A,B}\}} \Int_{\mathbb{R}^3}
      |\bra{_{\gamma}^1\Phi^{\vec k\I{P} \, \vec k\I{A}}_{ij}} \hat H_1
      \ket{^1\Phi^{\vec k\I{P}}_h}|^2 \\
    && \displaystyle\hspace{8em}{} \times \delta(E^{\vec k\I{P}}_h -
      E^{\vec k\I{P} \, \vec k\I{A}}_{ij,\gamma}) \differential^3 k\I{A} \\
    &\approx& \displaystyle 2\pi \Sum_{(i,j) \in \Cal F}
      \Int_{\mathbb{R}^3} 2 \, |v_{h\,\vec k\I{A}\,i\,j}|^2 \,
      \delta(E^{\vec k\I{P}}_h - E^{\vec k\I{P} \, \vec k\I{A}}_{ij})
      \differential^3 k\I{A} \; .
  \end{array}
\end{equation}
As we assume the two-step model of Auger decay~\cite{Aberg:EP-82,Aberg:UT-92},
$\Gamma_h$~is independent of the photoelectron momentum~$\vec k\I{P}$.
In Eqs.~(\ref{eq:resenshift}) and (\ref{eq:WWdecwidth}), we
mark explicitly the dependence on the hole orbital~$h \in \Cal H$.
As all~$h \in \Cal H$ are from a single magnetic subshell,
the~$\Delta_{\mathrm{R}, h}$ and the $\Gamma_h$ agree for all~$h \in \Cal H$.
Therefore, we may drop the dependence on~$h$ in what follows.

In the derivation of Eqs.~(\ref{eq:WWdecay}), (\ref{eq:resenshift}), and
(\ref{eq:WWdecwidth}), we implicitly assume that $\dot{\bar c}\I{P}^{\vec
k\I{P}}(t)$ varies only slightly on time intervals~$[t - \tau; t]$
for~$t \in [-\infty; \infty]$ and a small~$\tau > 0$ with
respect to all time scales in the problem~\cite{Yakovlev:ES-03,Buth:LA-08}.
Otherwise Eq.~(\ref{eq:WWdecay}) would not be meaningful.
We can integrate Eq.~(\ref{eq:EOMaugerOlgaFin}) formally,
\begin{equation}
  \label{eq:intAugerEss}
  \begin{array}{rcl}
    \displaystyle \bar c^{\vec k\I{P} \, \vec k\I{A}}_{ij}(t) &=&
      \displaystyle \Int_{t - \tau}^t \dot{\bar c}^{\vec k\I{P} \,
      \vec k\I{A}}_{ij}(t') \differential t' +
      \Int_{-\infty}^{t - \tau} \dot{\bar c}^{\vec k\I{P} \,
      \vec k\I{A}}_{ij}(t') \differential t' \\
    &=& \displaystyle \frac{1}{\sqrt{2}} \, \bar v(\vec k\I{A}) \>
      \bar c^{\vec k\I{P}}_h(t) \> \frac{\euler^{\imag (\frac{\vec
      k\I{A}^2}{2} - \Omega\I{A}- \imag \eta) t}}{\frac{\vec k\I{A}^2}{2}
      - \Omega\I{A} - \imag \eta} \; .
  \end{array}
\end{equation}
Here, $\euler^{\eta \, t}$ with~$\eta > 0$ ensures the initial
condition~$\bar c\I{A}^{\vec k\I{P} \, \vec k\I{A}} (-\infty) = 0$
and the convergence of the integral where $\eta \to 0^+$~is performed
after the integration.
The last equality follows from~$\Int^{t}_{t - \tau}
\dot{\bar c}\I{A}^{\vec k\I{P} \, \vec k\I{A}}(t') \differential t'
= \bar c\I{A}^{\vec k\I{P} \, \vec k\I{A}}(t) - \bar c\I{A}^{\vec
k\I{P} \, \vec k\I{A}}(t - \tau)$, where the term for~$t - \tau$
cancels the second integral.

Our result for the Auger amplitude~(\ref{eq:intAugerEss}) is inserted
into the expression for~$w(\vec k\I{P}, t)$
[the second term on the right-hand side of Eq.~(\ref{eq:EOMxuvOlgaFin})]
yielding
\begin{equation}
  w(\vec k\I{P}, t) = 2 \, \imag \, \Int_{\mathbb{R}^3}
    \frac{|\bar v(\vec k\I{A})|^2 \; \euler^{\eta \, t}}
    {\frac{\vec k\I{A}^2}{2} - \Omega\I{A} - \imag \eta} \;
    \bar c^{\vec k\I{P}}_h(t) \differential^3 k\I{A} \; .
\end{equation}
With the decomposition~\cite{Arfken:MM-05}
\begin{equation}
  \frac{1}{x - \imag \eta} = \Pr \frac{1}{x} + \imag \pi \delta(x) \; ,
\end{equation}
we obtain Eqs.~(\ref{eq:WWdecay}), (\ref{eq:resenshift}), and
(\ref{eq:WWdecwidth}) after dropping the subscript~$h$, eliminating the
sum over final states and replacing the energies and two-electron matrix
elements in Eqs.~(\ref{eq:resenshift}) and (\ref{eq:WWdecwidth}).
In detail, we find
\begin{equation}
  \label{eq:resenshiftAve}
  \begin{array}{rcl}
    \displaystyle \Delta_{\mathrm{R}} &=& \displaystyle
      \Pr \Int_{\mathbb{R}^3} \frac{2 \, |\bar v(\vec k\I{A})|^2}
      {\Omega\I{A} - \frac{\vec k\I{A}^2}{2}} \differential^3 k\I{A} \; ,
  \end{array}
\end{equation}
and
\begin{equation}
  \label{eq:WWdecwidthAve}
  \begin{array}{rcl}
    \displaystyle \Gamma &=& \displaystyle 2\pi \;
      \Int_{\mathbb{R}^3} 2 \, |\bar v(\vec k\I{A})|^2 \,
      \delta \bigl( \Omega\I{A} - \frac{\vec k\I{A}^2}{2} \bigr)
      \differential^3 k\I{A} \; .
  \end{array}
\end{equation}

A suitable value for the strength of the Auger decay matrix
element~(\ref{eq:AugME}) can be obtained from Eq.~(\ref{eq:WWdecwidthAve}) via
\begin{equation}
  \label{eq:strengthtwoel}
  Q_v = \sqrt{\dfrac{\Gamma\I{par}}{\Gamma |_{Q_v = 1}}} \; ,
\end{equation}
provided the decay width~$\Gamma\I{par}$ is taken to be an (experimental)
parameter.

\subsection{Hole-state amplitude with laser dressing}
\label{sec:hostamp}

In Secs.~\ref{sec:maelen} and \ref{sec:pertdec}, we
disregarded laser dressing and focused on the EOMs with \XUV{}~light only.
In the framework of Sec.~\ref{sec:laserdre}, we can easily incorporate
laser dressing in the strong-field approximation~\cite{Madsen:SF-05}
into our equations;
the only change in our EOMs~(\ref{eq:EOMxuvOlgaFin}) and
(\ref{eq:EOMaugerOlgaFin}) concerns the
time-dependent phase factors~$\imag \frac{\vec k\I{P}^2}{2} \, t$ which
need to be replaced by Volkov phases~$\imag \Phi\I{V}(\vec k\I{P}, t)$
[Eq.~(\ref{eq:VolkovPhaseCW})].
Using relation~(\ref{eq:WWdecay}), we decouple the
hole amplitude from the Auger decay amplitude and recast
Eq.~(\ref{eq:EOMxuvOlgaFin}) into
\begin{equation}
  \label{eq:EOMaugerOlgaWW}
  \begin{array}{rcl}
    \displaystyle \dot{\bar c}\I{P}^{\vec k\I{P}}(t) &=& \displaystyle
      \frac{-\imag}{\sqrt{2}} \> \bar d(\vec k\I{P}) \, \varepsilon\I{X}(t) \>
      \euler^{\imag \, \Phi\I{V}(\vec k\I{P}, t)} \>
      \euler^{-\imag \, \Omega\I{P} \, t} \\
    &&\displaystyle{} + \Bigl(-\imag \, \Delta\I{R} - \frac{\Gamma}{2} \Bigr)
      \> \bar c\I{P}^{\vec k\I{P}}(t) \; ,
  \end{array}
\end{equation}
assuming that the Auger decay is uninfluenced by the laser and thus
the second-order energy shift~$\Delta\I{R}$ [Eq.~(\ref{eq:resenshiftAve})]
and the Auger decay rate~$\Gamma$ [Eq.~(\ref{eq:WWdecwidthAve})]
are meaningful.
The first-order ordinary differential equation~(\ref{eq:EOMaugerOlgaWW})
is solved analytically~\cite{Arfken:MM-05} yielding
\begin{equation}
  \label{eq:EOMhalfInt}
  \begin{array}{rcl}
    \displaystyle \bar c\I{P}^{\vec k\I{P}}(t) &=& \displaystyle
      \frac{-\imag}{\sqrt{2}} \> \bar d(\vec k\I{P}) \,
      \euler^{-\imag \, (\Delta\I{R} - \imag \frac{\Gamma}{2}) \, t} \,
      \Int_{-\infty}^t \, \varepsilon\I{X}(t') \\
    &&\displaystyle{} \times \euler^{\imag \, \Phi\I{V}(\vec k\I{P}, t)} \;
      \euler^{\imag \, (\Delta\I{R} - \imag \frac{\Gamma}{2}
      - \Omega\I{P}) \, t'} \differential t' \; .
  \end{array}
\end{equation}

To solve the time integration in Eq.~(\ref{eq:EOMhalfInt}), we expand the
Volkov phase as in
Eq.~(\ref{eq:VolkovExpand}) and insert the inverse Fourier transform of
the envelope of the \XUV{}~pulse~(\ref{eq:XUVfield}),
\begin{equation}
  \label{eq:invFourier}
  \varepsilon\I{X}(t') = \frac{1}{2\pi} \Int^{\infty}_{-\infty}
    \tilde \varepsilon\I{X}(\omega) \, \euler^{-\imag \omega t'}
    \differential \omega \; .
\end{equation}
We obtain for the laser-dressed hole amplitude~(\ref{eq:EOMhalfInt}) the
expression
\begin{equation}
  \label{eq:EOMholeAmp}
  \begin{array}{rcl}
    \displaystyle \bar c\I{P}^{\vec k\I{P}}(t) &=& \displaystyle
      \frac{-1}{2 \, \sqrt{2} \, \pi} \> \bar d(\vec k\I{P})
      \Sum^{\infty}_{m=-\infty} \euler^{\imag \, m \, (\omega\I{L} \,
      \Delta t - \frac{\pi}{2})} \\
    &&\displaystyle{} \times J_m \Bigl( \vec \alpha\I{L} \mul \vec k\I{P},
      \frac{U\I{P}}{2 \, \omega\I{L}} \Bigr) \\
    &&\displaystyle{} \times \Int_{-\infty}^{\infty}
      \dfrac{\tilde \varepsilon\I{X}(\omega) \; \euler^{\imag \,
      (\frac{\vec k\I{P}^2}{2} + m \, \omega\I{L} + U\I{P} - \Omega\I{P}
      - \omega) \, t}} {\frac{\vec k\I{P}^2}{2} + m \, \omega\I{L}
      + U\I{P} - \Omega\I{P} + \Delta\I{R} - \omega
      - \imag \frac{\Gamma}{2}} \differential \omega \; .
  \end{array}
\end{equation}
For moderate laser intensities, we have~$\frac{U\I{P}}{2 \,
\omega\I{L}} \approx 0$.
Then, the generalized Bessel functions go over into ordinary Bessel
functions~\cite{Reiss:EI-80,Madsen:SF-05}.
Further, the limit~$\Lim_{u,v \to 0} J_m(u,v) = \delta_{m,0}$ exists
which completely removes the dependence of the equation on the laser
for vanishing intensity.
With this approximation, our expression~(\ref{eq:EOMholeAmp}) goes over into
Smirnova~\etal's~\cite{Smirnova:QC-03,footnote3} Eq.~(15).

\subsection{Laser-dressed Auger electron spectrum}
\label{sec:laser}

The amplitude to observe an Auger electron with momentum~$\vec k\I{A}$
for a photoelectron with momentum~$\vec k\I{P}$ at time~$t$ is found by
integrating Eq.~(\ref{eq:EOMaugerOlgaFin}) from~$-\infty$ to~$t$.
Beforehand, expression~(\ref{eq:EOMaugerOlgaFin}) needs to be adapted for laser
dressing by replacing~$\imag \, \frac{\vec k\I{A}^2}{2} \, t$
with~$\imag \, \Phi\I{V}(\vec k\I{A}, t)$
after which we insert the hole-state amplitude~(\ref{eq:EOMholeAmp}).
We obtain the following closed-form expression:
\begin{equation}
  \label{eq:AugerDressedIntermediate}
  \begin{array}{rcl}
    \displaystyle \bar c\I{A}^{\vec k\I{P} \, \vec k\I{A}}(t) &=&
      \displaystyle \frac{-\imag}{4\pi} \;
      \bar d(\vec k\I{P}) \> \bar v(\vec k\I{A}) \>
      \Sum^{\infty}_{m=-\infty} \euler^{\imag \, m \, (\omega\I{L} \,
      \Delta t - \frac{\pi}{2})} \\
    &&\displaystyle{} \times J_m \Bigl( \vec \alpha\I{L} \mul
      \vec k\I{P}, \frac{U\I{P}}{2 \, \omega\I{L}} \Bigr) \\
    &&\displaystyle{} \times
      \Int_{-\infty}^{\infty} \dfrac{\tilde \varepsilon\I{X}(\omega)}
      {\frac{\vec k\I{P}^2}{2} + m \, \omega\I{L}
      + U\I{P} - \Omega\I{P} + \Delta\I{R} - \omega - \imag \frac{\Gamma}{2}} \\
    &&\displaystyle{} \times \Int_{-\infty}^t \euler^{\imag
      \, ( \frac{\vec k\I{P}^2}{2} + m \, \omega\I{L} + U\I{P}
      - \Omega\I{P} - \Omega\I{A} - \omega ) \, t'} \\
    &&\displaystyle{} \times \euler^{\imag \, \Phi\I{V}(\vec k\I{A}, t')}
      \differential t' \differential \omega \; .
  \end{array}
\end{equation}
We are only interested in the Auger electron spectrum after
the \XUV{}~pulse is over and the induced hole amplitude has decayed.
Therefore, after expanding the Volkov phase~(\ref{eq:VolkovExpand}),
we let~$t \to \infty$ and simplify the time integration in
Eq.~(\ref{eq:AugerDressedIntermediate}) by observing that
\begin{equation}
  \label{eq:DiracFour}
  \delta(\omega - \omega') = \frac{1}{2\pi} \Int^{\infty}_{-\infty}
    \euler^{\imag \, (\omega - \omega') t'} \differential t'
\end{equation}
is a representation for Dirac's
$\delta$~distribution~\cite{Merzbacher:QM-98,Arfken:MM-05}.
Finally, we obtain the laser-dressed Auger amplitude,
\begin{equation}
  \label{eq:lasdreauampre}
  \begin{array}{rcl}
    \displaystyle \bar c\I{A}^{\vec k\I{P} \, \vec k\I{A}}(\infty) &=&
      \displaystyle \frac{\imag}{2} \>
      \bar d(\vec k\I{P}) \>
      \bar v(\vec k\I{A}) \> \Sum^{\infty}_{m,n = -\infty}
      \euler^{\imag \, (m+n) \, (\omega\I{L} \, \Delta t - \frac{\pi}{2})} \\
    &&\displaystyle{} \times
      J_m \Bigl( \vec \alpha\I{L} \mul \vec k\I{P}, \frac{U\I{P}}{2 \, \omega\I{L}} \Bigr) \;
      J_n \Bigl( \vec \alpha\I{L} \mul \vec k\I{A}, \frac{U\I{P}}{2 \, \omega\I{L}} \Bigr) \\
    &&\displaystyle{} \times
      S\Bigl( \frac{\vec k\I{P}^2}{2} + m \, \omega\I{L} + U\I{P},
              \frac{\vec k\I{A}^2}{2} + n \, \omega\I{L} + U\I{P} \Bigr) \; ,
  \end{array}
\end{equation}
with the line shape function
\begin{equation}
  \label{eq:lineshape}
  S(\omega\I{P}, \omega\I{A}) = \frac{\tilde \varepsilon\I{X}
    (\omega\I{P} + \omega\I{A} - \Omega\I{P} - \Omega\I{A})}
    {\omega\I{A} - \Omega\I{A} - \Delta\I{R} + \imag \frac{\Gamma}{2}} \; ,
\end{equation}
which depends only on the absolute values of the momenta~$\vec k\I{P}$
and $\vec k\I{A}$.
Formula~(\ref{eq:lasdreauampre}) goes over into Eq.~(18) in
Ref.~\onlinecite{Smirnova:QC-03}---apart from a factor~$\frac{\imag}{2}$
in our expression---by setting~$U\I{P} = 0$, replacing the generalized
Bessel functions by ordinary ones and removing the dependence on the
laser for the photoelectrons by using~$J_m(0) = \delta_{m,0}$.

We are now in a position to determine the laser-dressed Auger
electron spectrum where we consider the case that the photoelectron
is not observed.
Therefore, we integrate the probability density~$|\bar c\I{A}^{\vec k\I{P} \,
\vec k\I{A}}(\infty)|^2$ [Eqs.~(\ref{eq:lasdreauampre}) and
(\ref{eq:lineshape})] over all possible photoelectron momenta to
eliminate this degree of freedom.
This yields for the probability density~\cite{Merzbacher:QM-98}
to observe an Auger electron with momentum vector~$\vec k\I{A}$,
\begin{equation}
  \label{eq:AugElSpec}
  P\I{A}(\vec k\I{A}) = \Int_{\mathbb{R}^3} |\bar c\I{A}^{\vec k\I{P}
    \, \vec k\I{A}}(\infty)|^2 \differential^3 k\I{P} \; .
\end{equation}

\subsection{\XUV{}~absorption cross section of laser-dressed atoms}
\label{sec:xsect}

The probability of finding an atom in the ground state is given in terms
of the ground-state amplitude~$\bar c_0(t)$ in the wave
packet~(\ref{eq:wavepacket}) by
\begin{equation}
  \Cal P_0(t) = |\bar c_0(t)|^2 = \bar c_0^*(t) \,
    \bar c_0(t) \; .
\end{equation}
Consequently, the negative of the \XUV{}-absorption rate~\cite{Buth:TX-07} is
\begin{eqnarray}
  \label{eq:XUVrate}
  -\Gamma\I{X}(t) = \dot{\Cal P}_0(t) &=& \dot{\bar c}_0^*(t) \,
    \bar c_0(t) + \bar c_0^*(t) \, \dot{\bar c}_0(t)
    \nonumber \\
  &\approx& \dot{\bar c}_0^*(t) + \dot{\bar c}_0(t) \\
  &=& 2 \  \re \dot{\bar c}_0(t)
   =  2 \  \im [\imag \> \dot{\bar c}_0(t)] \; . \nonumber
\end{eqnarray}
The center line follows from the weakness of \XUV{}~absorption,
\ie, $\bar c_0(t) \approx 1$ for all~$t$.
The rate of change of the ground-state amplitude follows from
the first EOM~(\ref{eq:EOMground});
adapted for the essential-states model with laser dressing, it reads
\begin{equation}
  \label{eq:EOMgroundLD}
  \dot{\bar c}_0(t) = \frac{-\imag}{\sqrt{2}}
    \Int_{\mathbb R^3} \bar d^*(\vec k\I{P}) \, \varepsilon\I{X}(t) \,
    \euler^{\imag \, \Omega\I{P} \, t} \,
    \euler^{-\imag \, \Phi\I{V}(\vec k\I{P}, t)} \,
    \bar c\I{P}^{\vec k\I{P}}(t) \differential^3 k\I{P} \; .
\end{equation}
Inserting this EOM into Eq.~(\ref{eq:XUVrate}),
expanding the Volkov phase factor using Eq.~(\ref{eq:VolkovExpand}), and
inserting the laser-dressed hole-state amplitude~(\ref{eq:EOMholeAmp}),
we obtain the rate
\begin{widetext}
\begin{equation}
  \label{eq:xuvratexsect}
  \begin{array}{rcl}
    \displaystyle \Gamma\I{X}(t) &=& \displaystyle -2 \  \im [\imag \>
      \dot{\bar c}_0(t)] = \frac{1
        }{2 \pi} \, \varepsilon\I{X}(t)
      \Sum_{m, n = -\infty}^{\infty} \  \Int_{\mathbb R^3}
      J_m \Bigl( \vec \alpha\I{L} \mul \vec k\I{P},
      \frac{U\I{P}}{2 \, \omega\I{L}} \Bigr) \
      J_n \Bigl( \vec \alpha\I{L} \mul \vec k\I{P},
      \frac{U\I{P}}{2 \, \omega\I{L}} \Bigr) \\
    &&\displaystyle{} \times |\bar d(\vec k\I{P})|^2 \  \Int_{-\infty}^{\infty}
      \  \im \biggl [ \dfrac{ \tilde \varepsilon\I{X}(\omega) \;
      \euler^{-\imag \, \omega \, t} \; \euler^{\imag \, (n-m) \,
      [\omega\I{L} \, (t + \Delta t) - \frac{\pi}{2}]}}
      {\frac{\vec k\I{P}^2}{2} + n \, \omega\I{L} + U\I{P} - \Omega\I{P}
      + \Delta\I{R} - \omega - \imag \frac{\Gamma}{2}}
      \biggr ] \differential \omega \differential^3 k\I{P} \; .
  \end{array}
\end{equation}

The absorption rate~(\ref{eq:xuvratexsect}) in conjunction with
the flux~$J\I{X} = \frac{I\I{X,0}}{\omega\I{X}}$ at the
\XUV{}~(peak) intensity~$I\I{X,0}$ with photon energy~$\omega\I{X}$
allows one to obtain the \XUV{}~photoabsorption cross
section~\cite{footnote5} via
\begin{equation}
  \label{eq:xsect}
  \sigma = \frac{\Gamma\I{X}}{J\I{X}} \; .
\end{equation}
Note that for a continuous-wave approximation of monochromatic
radiation, we have for the \XUV{}~field
strength~$\tilde \varepsilon\I{X}(\omega) = \tilde \varepsilon\I{X,0}
\> \delta(\omega)$ and $\varepsilon\I{X}(t) = E\I{X,0}
= \sqrt{8 \pi \, \alpha \, I\I{X,0}} = \hbox{const}$ for all~$t$.
The relation between time- and frequency-domain field amplitudes
follows from Eq.~(\ref{eq:invFourier}) and
is~$E\I{X,0} = \frac{1}{2\pi} \, \tilde \varepsilon\I{X,0}$.
For large~$t$, terms with~$m \neq n$ oscillate rapidly
in the absorption rate~(\ref{eq:xuvratexsect}).
We discard these terms and retain only the constant terms with~$m = n$.
This leads to the expression
\begin{equation}
  \label{eq:xsectXUV}
  \sigma(\omega\I{X}) = 8 \pi \, \alpha \, \omega\I{X}
    \Sum_{m = -\infty}^{\infty} \  \Int_{\mathbb R^3}
    J_m \Bigl( \vec \alpha\I{L} \mul \vec k\I{P},
    \frac{U\I{P}}{2 \, \omega\I{L}} \Bigr)^2
    \  \im \Bigl [ \dfrac{|\bar d(\vec k\I{P})|^2}
    {E^+ + \frac{\vec k\I{P}^2}{2} + m \, \omega\I{L} + U\I{P} - E_0
    + \Delta\I{R} - \omega\I{X} - \imag \frac{\Gamma}{2}}
    \Bigr ] \differential^3 k\I{P} \; ,
\end{equation}
\end{widetext}
by expanding~$\Omega\I{P} = E_0 - E^+ + \omega\I{X}$.
For vanishing laser intensity, the structure of this equations
becomes the same as from Eq.~(40) in Ref.~\onlinecite{Buth:TX-07}.
There, however, $\Delta\I{R}$ was not accounted for.

Similarly to Eq.~(\ref{eq:strengthtwoel}), a suitable value for the
strength of the dipole matrix element~(\ref{eq:DipME}) can be obtained
from Eq.~(\ref{eq:xsectXUV}), omitting laser dressing, via
\begin{equation}
  \label{eq:strengthdipole}
  Q_d = \sqrt{\dfrac{\sigma\I{par}(\omega\I{par})}{\sigma(\omega\I{par})
    | _{Q_d = 1}}} \; ,
\end{equation}
provided the cross section~$\sigma\I{par}(\omega\I{par})$ at an
energy~$\omega\I{par}$ in the range of energies of interest or close to
the range is taken as a (experimental) parameter.

\subsection{Laser-dressed photoelectron spectrum}
\label{sec:ldpel}

We determined the rate with which \XUV{}~light is
absorbed~$\Gamma\I{X}(t)$ [Eq.~(\ref{eq:XUVrate})] in Sec.~\ref{sec:xsect}.
The rate was derived under the premise of weak \XUV{}~absorption
which allowed us to approximate the ground-state amplitude
by~$\bar c_0(t) \approx 1$ for all~$t$.
Therefore, $|\bar c_0(t)|^2$~cannot be used
to obtain the probability with which photoelectrons are ejected.
Instead, we need to integrate the rate
\begin{equation}
  \label{eq:photoabs}
  \Cal P\I{P}(t) = \Int_{-\infty}^t \Gamma\I{X}(t') \differential t'
    = -2 \Int_{-\infty}^t \im [\imag \> \dot{\bar c}_0(t')]
    \differential t' \; .
\end{equation}
In expression~(\ref{eq:photoabs}), we insert the ground-state amplitude rate of
change~(\ref{eq:EOMgroundLD}) and the laser-dressed hole-state
amplitude~(\ref{eq:EOMholeAmp}) and expand the Volkov phase
factor~(\ref{eq:VolkovExpand}) to obtain the probability
\begin{widetext}
\begin{equation}
  \begin{array}{rcl}
    \displaystyle \Cal P\I{P}(t) &=& \displaystyle \frac{1
        }{2 \pi}
      \Sum_{m, n = -\infty}^{\infty} \Int_{\mathbb R^3}
      J_m \Bigl( \vec \alpha\I{L} \mul \vec k\I{P},
      \frac{U\I{P}}{2 \, \omega\I{L}} \Bigr) \;
      J_n \Bigl( \vec \alpha\I{L} \mul \vec k\I{P},
      \frac{U\I{P}}{2 \, \omega\I{L}} \Bigr)
      \  \Int_{-\infty}^t  \varepsilon\I{X}(t')
      \  \Int_{-\infty}^{\infty} \  \im \biggl [
      \tilde \varepsilon\I{X}(\omega) \\
    &&\displaystyle{} \times \dfrac{|\bar d(\vec k\I{P})|^2 \
      \euler^{-\imag \, \omega \, t} \, \euler^{\imag \, (n-m) \,
      [\omega\I{L} \, (t + \Delta t) - \frac{\pi}{2}]}}
      {\frac{\vec k\I{P}^2}{2} + n \, \omega\I{L} + U\I{P} - \Omega\I{P}
      + \Delta\I{R} - \omega - \imag \frac{\Gamma}{2}}
      \biggr ] \differential \omega \differential t'
      \differential^3 k\I{P} \; .
  \end{array}
\end{equation}
Letting~$t \to \infty$, replacing the real-valued~$\varepsilon\I{X}(t')$
by the complex conjugate of
Eq.~(\ref{eq:invFourier}), and omitting the integration over~$\vec k\I{P}$,
we find the probability density for photoelectron ejection by the
\XUV{}~pulse.
The time integration yields a $\delta$~distribution~(\ref{eq:DiracFour})
of the form~$\delta(\omega' - \omega + (n-m) \, \omega\I{L})$.
Replacing $\tilde m \equiv n - m$, we arrive at the probability
density
\begin{equation}
  \label{eq:pelspec}
  \begin{array}{rcl}
    \displaystyle \tilde{\Cal P}\I{P}(\vec k\I{P}) &\equiv&
      \displaystyle \frac{1}{2 \pi} \; |\bar d(\vec k\I{P})|^2
      \Sum_{\tilde m = -\infty}^{\infty} \Sum_{n = -\infty}^{\infty}
      J_{n - \tilde m} \Bigl( \vec \alpha\I{L} \mul \vec k\I{P},
      \frac{U\I{P}}{2 \, \omega\I{L}} \Bigr) \;
      J_n \Bigl( \vec \alpha\I{L} \mul \vec k\I{P},
      \frac{U\I{P}}{2 \, \omega\I{L}} \Bigr)   \\
    &&\displaystyle{} \times \im \biggl [ \Int_{-\infty}^{\infty}
      \dfrac{ \euler^{\imag \, \tilde m \, [\omega\I{L} \, \Delta t
      - \frac{\pi}{2}]} \; \tilde \varepsilon\I{X}^*(\omega
      - \tilde m \, \omega\I{L}) \; \tilde \varepsilon\I{X}(\omega)}
      {\frac{\vec k\I{P}^2}{2} + n \, \omega\I{L} + U\I{P} - \Omega\I{P}
      + \Delta\I{R} - \omega - \imag \frac{\Gamma}{2}}
      \differential \omega \biggr ] \; .
  \end{array}
\end{equation}
\end{widetext}
The photoelectron spectrum depends on the Fourier transform of the
\XUV{}~field envelope at~$\omega$ and at~$\omega - \tilde m \, \omega\I{L}$.
This functional dependence indicates interference effects between
channels with a different number of laser photons provided that
the \XUV{}~field envelope has sufficient width.

\section{Electronic structure}
\label{sec:elstructmodel}

The theory of Secs.~\ref{sec:auger}, \ref{sec:laserdre}, and \ref{sec:essstmo}
treated the electronic structure of an atom as an abstract quantity
which was represented by the orbital energies in~$\hat H\I{HFS}$
[Eq.~(\ref{eq:H_HFS})], the one-electron matrix elements
in~$\hat H\I{CH}$ [Eq.~(\ref{eq:H_CH})], the dipole matrix
elements in~$\hat H\I{X}$ [Eq.~(\ref{eq:xuvint})],
and the two-electron matrix elements in~$\hat H\I{ee}$ [Eq.~(\ref{eq:Hee})].
Programs exist to carry out the Hartree-Fock-Slater
approximation~\cite{Slater:AS-51,Slater:XA-72}
and compute the required one- and two-electron matrix elements.
To evaluate the essential-states model of Sec.~\ref{sec:essstmo},
however, we use a much simpler model approach in terms of scaled
hydrogenic functions for the atomic orbitals.
This treatment follows
Refs.~\onlinecite{Yakovlev:TR-02,Yakovlev:ES-03,Smirnova:QC-03}.
The parametrization of the model corrects to a large extend for
inaccuracies in the orbital energies and the matrix elements.
If the results of the essential-states model depended sensitively
on the electronic structure, then due to the substantial simplifications made,
its physical predictions would be untrustworthy.
Despite the use of approximate orbitals,
the equations derived in this section are completely
general and an \emph{ab initio} evaluation in terms
of Hartree-Fock-Slater orbitals is feasible.

We use hydrogenic wave functions to model the spatial atomic
orbitals in spherical polar coordinates~$\tilde \varphi_i(r, \vartheta,
\varphi) = R_{n_i \, l_i}(r) \> Y_{l_i \, m_i}(\vartheta,
\varphi)$~\cite{Merzbacher:QM-98}.
Here, $n_i$, $l_i$, and $m_i$ are the principal, orbital angular momentum, and
magnetic quantum number, respectively, of orbital~$i \in \{1, \ldots, Z/2\}$.
The radial part is~$R_{n_i \, l_i}(r)$ and the angular dependence is
described by spherical harmonics~$Y_{l_i \, m_i}(\vartheta, \varphi)$.
We scale the hydrogenic wave functions such that
their energy~$E_{n_i} = -\frac{Z^2}{2 n_i^2}$ matches
the energy~$\varepsilon_i$ of the corresponding
orbital in the chosen atom~\cite{Yakovlev:TR-02,Yakovlev:ES-03}.
For this purpose we use an effective charge
\begin{equation}
  \label{eq:effcharge}
  Z_{\mathrm{eff}, i} = n_i \sqrt{-2 \, \varepsilon_i} \; .
\end{equation}
This scaling also adjusts the spatial extend of the orbital appropriately.

\subsection{Dipole matrix elements}

The dipole matrix elements for \XUV{}~absorption~(\ref{eq:XUVdipole}) are
given by the promoted wave function in momentum
space~\cite{Talukdar:RM-84,Merzbacher:QM-98},
\begin{equation}
  \label{eq:XUVdipPlane}
  d_{\vec k\I{P} \, h}
    = \frac{1}{(2\pi)^{3/2}}
    \Int_{\mathbb R^3} \euler^{-\imag \, \vec k \mul \vec r} \>
    z \; \varphi_h(\vec r) \differential^3 r \; ,
\end{equation}
with~$\vec r \mul \vec e\I{X} = z$.
We use an atomic orbital~$\varphi_h(\vec r)$ for the vacancy
created by photoionization and the spatial part~$\varphi_{\mathrm P,
\vec k}(\vec r, 0)$ of a plane wave~(\ref{eq:planewave}).
This comprises also the case when laser dressing is considered because
the laser dressing manifests itself exclusively~\cite{footnote1} in the
time-dependent Volkov phase~(\ref{eq:VolkovPhase}).
We use the Rayleigh expansion~\cite{Rose:ET-57} of the plane waves
in Eq.~(\ref{eq:XUVdipPlane}),
\begin{equation}
  \label{eq:sphereexpand}
  \euler^{\imag \vec k \mul \vec r} = 4\pi \Sum_{l=0}^{\infty} \Sum_{m=-l}^l
    \imag^l \, Y^*_{lm}(\vartheta_k, \varphi_k) \, j_l(k \, r)
    \, Y_{lm}(\vartheta, \varphi) \; .
\end{equation}
The directions of~$\vec k$ and $\vec r$ are specified by the polar
angles~$\vartheta_k$, $\varphi_k$ and $\vartheta$, $\varphi$, respectively.
Here, $j_l$~denotes a spherical Bessel function~\cite{Arfken:MM-05}.
We arrive at the dipole matrix element~(\ref{eq:XUVdipPlane})
in spherical polar coordinates,
\begin{widetext}
\begin{equation}
  \label{eq:XUVdipexpanded}
  \tilde d_h(k\I{P}, \vartheta_{k\I{P}}, \varphi_{k\I{P}}) = 2 \,
    \sqrt{\frac{2}{3}} \Sum_{\atopa{l \in \{l_h-1, l_h+1\}}{l \geq 0}}
    (-\imag)^l \> \threeY{l_h,1,l}{m_h,0,m_h} \> Y_{l \, m_h}
    (\vartheta_{k\I{P}}, \varphi_{k\I{P}}) \,
    \Cal D_{n_h \, l_h}^{(l)}(k\I{P}) \; .
\end{equation}
Corresponding to orbital~$h$, we have the principal~$n_h$, orbital
angular momentum~$l_h$, and magnetic~$m_h$ quantum numbers.
The angular integral is
\begin{equation}
  \label{eq:threeY}
  \begin{array}{rcl}
    \threeY{l_1,l_2,l_3}{m_1,m_2,m_3} &=& \Int_{4\pi} Y^*_{l_3 \, m_3}(\Omega)
      \, Y_{l_2 \, m_2}(\Omega) \, Y_{l_1 \, m_1}(\Omega) \differential
      \Omega \\
    &=& \sqrt{\frac{(2 l_1 + 1) (2 l_2 + 1)}{4 \pi \, (2 l_3 + 1)}} \>
      \cleb{l_1,l_2,l_3}{m_1,m_2,m_3} \> \cleb{l_1,l_2,l_3}{0,0,0} \; ,
  \end{array}
\end{equation}
\end{widetext}
where $\cleb{l_1,l_2,l_3}{m_1,m_2,m_3}$~is a Clebsch-Gordan
coefficient~\cite{Rose:ET-57}.
The integral restricts the accessible angular momenta and magnetic
quantum numbers in the photoionization process.
The radial dipole matrix elements are
\begin{equation}
  \Cal D_{n_h \, l_h}^{(l)}(k\I{P}) = \Int_0^{\infty} j_l(k\I{P} \, r) \,
    r^3 \, R_{n_h \, l_h}(r) \differential r \; ,
\end{equation}
in terms of the radial part~$R_{n_h \, l_h}(r)$ of the atomic orbital~$h$.

\subsection{Auger transition matrix elements}

Auger decay is mediated by the two-electron matrix element,
\begin{equation}
  \label{eq:twoelAuger}
  v_{h \, \vec k\I{A} \, ij} \equiv \tilde \chi_h^{ij}(\vec k\I{A})
    = \frac{1}{(2\pi)^{3/2}} \Int_{\mathbb{R}^3} \euler^{-\imag
    \vec k\I{A} \mul \vec r} \, \chi_h^{ij}(\vec r) \differential^3 r \; .
\end{equation}
Here, $\chi_h^{ij}(\vec r)$ and $\tilde \chi_h^{ij}(\vec k\I{A})$~are the
configuration space and momentum space Auger electron wave
functions, respectively~\cite{Merzbacher:QM-98,Yakovlev:TR-02}.
With the two-electron repulsion~$\hat h\I{ee}$ [Eq.~(\ref{eq:twoelCoulomb})],
the configuration space Auger electron wave functions reads
\begin{equation}
  \label{eq:AugerWaveConfig}
  \chi_h^{ij}(\vec r) =
    \varphi_j  (\vec r) \, \Int_0^{\infty} \Int_{4\pi}
    \tilde \varphi_h^*(r', \Omega') \; \hat h\I{ee} \;
    \tilde \varphi_i  (r', \Omega') \, r'^2 \differential r'
    \differential \Omega' \; .
\end{equation}
To simplify~$\hat h\I{ee}$, we replace it by the Laplace
expansion~\cite{Jackson:CE-98},
\begin{equation}
  \frac{1}{|\vec r - \vec r'|} = \Sum_{l=0}^{\infty}
    \frac{4\pi}{2l + 1} \, \varrho_l(r, r') \Sum_{m=-l}^l
    Y^*_{lm}(\vartheta', \varphi') \, Y_{lm}(\vartheta, \varphi) \; ,
\end{equation}
with the decomposition
\begin{equation}
  \label{eq:radialdep}
  \varrho_l(r, r') = \frac{r'^l}{r^{l+1}} \> \theta(r - r') +
                     \frac{r^l}{r'^{l+1}} \> \theta(r' - r) \; ,
\end{equation}
for the radial dependence, where $\theta$~is the Heaviside step function
with~$\theta(0) = \frac{1}{2}$.
The wave function of the Auger electron is in spherical polar coordinates,
\begin{equation}
  \label{eq:AugerWaveConfigDecomp}
  \begin{array}{rcl}
    \displaystyle \check \chi_h^{ij}(r, \vartheta, \varphi) &=& \displaystyle
      \tilde \varphi_j(r, \vartheta, \varphi) \Sum_{\atopa{l
      \mathrm{\; with\; } \Delta(l_h \, l \, l_i)}{l_i+l_h+l
      \mathrm{\; even}}} \dfrac{4\pi}{2 \, l + 1} \\
    &&\displaystyle{} \times \threeY{l_h,l,l_i}{m_h,m_i-m_h,m_i} \\
    &&\displaystyle{} \times \Cal R^{(1)}_{n_h\,l_h, n_i\,l_i, l}(r)
      \, Y_{l \, m_i-m_h}(\vartheta, \varphi) \; .
  \end{array}
\end{equation}
The symbol~$\Delta(l_h \, l \, l_i)$ represents the triangular condition
for which Clebsch-Gordan coefficients do not vanish~\cite{Rose:ET-57}.
The radial dependence in Eq.~(\ref{eq:AugerWaveConfigDecomp}) is
expressed by
\begin{equation}
  \label{eq:oneDrad}
  \Cal R^{(1)}_{n_h\,l_h, n_i\,l_i, l}(r) = \Int_0^{\infty}
    R_{n_h \, l_h}(r') \, \varrho_l(r, r') \, R_{n_i \, l_i}(r') \,
    r'^2 \differential r' \; .
\end{equation}

We would like to calculate the momentum space
representation~(\ref{eq:twoelAuger}) of the Auger electron wave
function~(\ref{eq:AugerWaveConfigDecomp})~\cite{Talukdar:RM-84,%
Merzbacher:QM-98}.
The plane wave is expanded in terms of spherical Bessel
functions~(\ref{eq:sphereexpand}),
\begin{equation}
  \label{eq:twopwexpand}
  \begin{array}{rcl}
    &&\displaystyle \acute \chi_h^{ij}(k\I{A}, \vartheta_{k\I{A}},
      \varphi_{k\I{A}}) \\
    &=& \displaystyle 4 \sqrt{2\pi}
      \Sum_{\atopa{l \mathrm{\; with\; } \Delta(l_h \, l \, l_i)}
      {l_i+l_h+l \mathrm{\; even}}} \dfrac{1}{2 \, l + 1} \\
    &&\displaystyle{} \times \threeY{l_h,l,l_i}{m_h,m_i-m_h,m_i}
      \, \Sum_{\atopa{l' \mathrm{\; with\; } \Delta(l \, l' \, l_j)}
      {l+l'+l_j \mathrm{\; even}}} (-\imag)^{l'} \\
    &&\displaystyle{} \times \threeY{l,l_j,l'}{m_i-m_h,m_j,m_i+m_j-m_h} \\
    &&\displaystyle{} \times \Cal R^{(2)}_{n_h\,l_h, n_i\,l_i, n_j\,l_j,
      l, l'}(k\I{A}) \, Y_{l' \, m_i+m_j-m_h}(\vartheta_{k\I{A}},
      \varphi_{k\I{A}})
  \end{array}
\end{equation}
with
\begin{equation}
  \label{eq:twoDrad}
  \begin{array}{rcl}
    \displaystyle \Cal R^{(2)}_{n_h\,l_h, n_i\,l_i, n_j\,l_j, l, l'}(k\I{A})
      &=& \displaystyle \Int_0^{\infty} j_{l'}(k\I{A} \, r) \,
      \Cal R^{(1)}_{n_h\,l_h, n_i\,l_i, l}(r) \\
    &&\displaystyle{} \times R_{n_j \, l_j}(r) \, r^2 \differential r \; .
  \end{array}
\end{equation}

We can decompose~$\Cal R^{(2)}_{n_h\,l_h, n_i\,l_i, n_j\,l_j, l,
l'}(k\I{A})$~(\ref{eq:twoDrad}) into a product of two independent
one-dimensional integrals by assuming the product ansatz
\begin{equation}
  \label{eq:radiusprod}
  \varrho_l(r, r') = \varrho_{1,l}(r) \, \varrho_{2,l}(r')
\end{equation}
for the radial dependence~(\ref{eq:radialdep}).
This simplifies our task to evaluate Eq.~(\ref{eq:twoDrad}) greatly.
It becomes
\begin{equation}
  \label{eq:radparttwo}
  \Cal R^{(2)}_{n_h\,l_h, n_i\,l_i, n_j\,l_j, l, l'}(k\I{A}) =
    \Cal R^{(3)}_{n_j\,l_j, l, l'}(k\I{A}) \,
    \Cal R^{(4)}_{n_h\,l_h, n_i\,l_i, l} \; ,
\end{equation}
with
\begin{equation}
  \Cal R^{(3)}_{n_j\,l_j, l, l'}(k\I{A}) = \Int_0^{\infty}
    j_{l'}(k\I{A} \, r) \, \varrho_{1,l}(r) \,
    R_{n_j \, l_j}(r) \, r^2 \differential r
\end{equation}
and
\begin{equation}
  \Cal R^{(4)}_{n_h\,l_h, n_i\,l_i, l} = \Int_0^{\infty}
    R_{n_h \, l_h}(r') \, \varrho_{2,l}(r') \,
    R_{n_i \, l_i}(r') \, r'^2 \differential r' \; .
\end{equation}

\section{Computational details}
\label{sec:compdet}

All computations were carried out with
\textsc{mathematica}~\cite{Mathematica:pgm-V7.0}.
In our essential-states model of Auger decay, the set of hole
orbitals~$\Cal H$ comprises all five~$3d$~orbitals of krypton.
The set of final states~$\Cal F$ consists of pairs of orbitals, the first
is the $4s$~orbital and the second is a $4p$~orbital with magnetic
quantum number~$m \in \{-1, 0, 1\}$.
The atomic orbital energies of krypton are taken from
Ref.~\onlinecite{Yakovlev:ES-03}:
$\varepsilon_{3d} = -70 \eV$, $\varepsilon_{4s} = -15 \eV$, and
$\varepsilon_{4p} = -15 \eV$.
The effective charges~(\ref{eq:effcharge}) assume the
%
%
values~$Z_{3d} = 6.8$, $Z_{4s} = 4.2$, and $Z_{4p} = 4.2$.
The orbital energies lead us to state energies (disregarding~$\hat H\I{CH}
+ \hat H\I{ee}$) via Eqs.~(\ref{eq:groundenergy}),
(\ref{eq:singlyenergy}), and (\ref{eq:doublyenergy}).
This yields, for \XUV{}~photons with~$\omega\I{X} = 90 \eV$, a nominal
photoelectron energy of~$\Omega\I{P} = E_0 - E^+ + \omega\I{X}
= \varepsilon_{3d} + \omega\I{X} = 20
\eV$~\cite{Yakovlev:ES-03,Smirnova:QC-03}.
We obtain a nominal Auger electron energy of~$\Omega\I{A} = E^+ - E^{++}
= \varepsilon_{4s} + \varepsilon_{4p} - \varepsilon_{3d} = 40 \eV$.
The Auger decay width of a $3d$~hole in krypton is \emph{artificially} set
%
%
to~$\Gamma\I{broad} = 1.3 \eV$, which corresponds to a decay time
of~$500 \U{as}$ in accord with the data in Fig.~3 in
Ref.~\onlinecite{Smirnova:QC-03}.
We use this much shorter decay time to show the coherence
in the laser-dressed Auger spectrum.
The experimental value for the decay width
%
%
is~$\Gamma\I{expt} = 88 \U{meV}$ which corresponds to a decay
time of~$7.5 \U{fs}$~\cite{Jurvansuu:IL-01}.

With an approximation which we discuss below [see Eq.~(\ref{eq:onemulti})]
for the radial dependence~(\ref{eq:radiusprod}), we determine the strengths
of the dipole~(\ref{eq:DipME}) and Auger decay~(\ref{eq:AugME}) matrix
elements.
Using the decay width~$\Gamma |_{Q_v = 1}$ from
Eq.~(\ref{eq:WWdecwidthAve}), we find from
Eq.~(\ref{eq:strengthtwoel}) the strengths~%
%
%
$Q_{v,\mathrm{expt}} = 1.10$ with~$\Gamma\I{expt}$ and
%
%
$Q_{v,\mathrm{broad}} = 4.25$ with~$\Gamma\I{broad}$.
The corresponding energy shifts follow from Eq.~(\ref{eq:resenshiftAve}).
They
%
%
are~$\Delta_{\mathrm{R,  expt}} = -0.90 \eV$ for~$Q_{v,\mathrm{expt}}$
%
%
and $\Delta_{\mathrm{R, broad}} = -13.53 \eV$ for~$Q_{v,\mathrm{broad}}$.
For a good agreement with the reference data [see Fig.~\ref{fig:AugerFig3}
below], we employ the
%
%
%
value~$Q_v = 3.1$ and the
%
%
shift~$\Delta_{\mathrm R} = -0.68 \eV$.
However, we set~$\Delta_{\mathrm{R}} = 0$ in all our computations because
we have chosen the orbital energies such that they correctly
reproduce the (experimental) Auger and photoelectron energies.
%
%
%
%
The photoionization cross section of the krypton $3d$~subshell
for~$20 \eV$~photoelectron energy in Hartree-Fock-Slater
approximation~\cite{footnote4}
is read off of graph~I in Ref.~\onlinecite{Yeh:AS-85};
it is about~$\sigma\I{HFS} = 1.5 \U{Mbarn}$.
The cross section without laser dressing~$\sigma(\omega\I{X})|_{Q_d = 1}$
is determined from Eq.~(\ref{eq:xsectXUV}) using~$\Gamma\I{expt}$
and letting~$\Delta_{\mathrm{R}} = 0$.
With Eq.~(\ref{eq:strengthdipole}), we obtain a dipole strength
%
%
of~$Q_d = 0.26$.

We use a \XUV{}~light pulse which has a Gaussian envelope with peak
intensity~$I\I{X,0}$ at~$t = 0$ and a full width at half maximum
duration of~$\tau\I{X} = 500 \U{as}$
\begin{equation}
  \label{eq:GaussEnv}
  I\I{X}(t) = I\I{X,0} \> \euler^{-4 \ln 2 (\frac{t}{\tau\I{X}})^2} \; .
\end{equation}
The \XUV{}~electric-field envelope in Eq.~(\ref{eq:XUVfield})
follows from~$\epsilon\I{X}(t) = \sqrt{8 \pi \, \alpha \, I\I{X}(t)}$;
its Fourier transform is also a Gaussian~\cite{Arfken:MM-05}.

\section{Results and discussion}
\label{sec:results}

We devote this section to a computational study of our essential-states
model and its parameters applied to $M_{4,5} N_1 N_{2,3}$~Auger decay in
krypton~\cite{Aksela:CE-84,Carlson:AD-89,Jurvansuu:IL-01,Schmidtke:AD-01}.
It is motivated by a previous experiment~\cite{Drescher:TR-02}
which focused on the line group around~$40 \eV$ and related theoretical
studies~\cite{Yakovlev:TR-02,Smirnova:QC-03,Yakovlev:ES-03}.
We assess the accuracy of the approximations made in our essential-states
model and compare with existing literature results.
In future work~\cite{Buth:AD-09}, we will explore laser-dressed
Auger decay extensively.
Additionally, we present the laser-dressed Auger spectrum for a much
higher dressing-laser intensity than what has been used so far.

\begin{figure}
  \begin{center}
    \includegraphics[clip,width=\hsize]{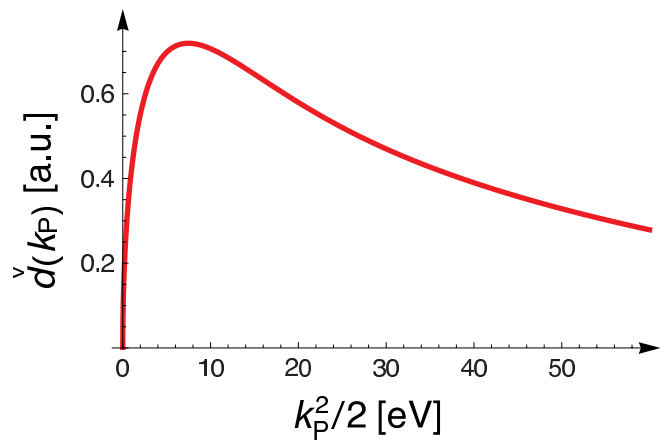}
    \caption{(Color online) Spherically integrated rms dipole matrix
             element~$\check d(k\I{P})$ [Eq.~(\ref{eq:avediplmodel})]
             for the description of $3d$~photoionization of a krypton atom
             in the essential-states model of Sec.~\ref{sec:essstmo}.}
    \label{fig:AugerDipole}
  \end{center}
\end{figure}

To begin with, let us discuss the results for the rms~matrix
elements~(\ref{eq:DipME}) and (\ref{eq:AugME}) of the essential-states model.
First, \XUV{}~absorption is determined by the rms dipole matrix element
in spherical polar coordinates~(\ref{eq:DipME}) which is determined from
the~$\tilde d_h(k\I{P}, \vartheta_{k\I{P}}, \varphi_{k\I{P}})$
[Eq.~(\ref{eq:XUVdipexpanded})] for all~$h \in \Cal H$.
We take its modulus squared and integrate over the full solid angle
to obtain the spherically integrated rms dipole matrix element,
\begin{equation}
  \label{eq:avediplmodel}
  \check d(k\I{P}) = \sqrt{\Int_{4\pi} |\tilde d(k\I{P},
    \Omega_{k\I{P}})|^2 \differential \Omega_{k\I{P}}} \; .
\end{equation}
It is plotted in Fig.~\ref{fig:AugerDipole}.
After a steep rise at the edge (zero photoelectron momentum),
it decays smoothly.
Around the nominal photoelectron energy of~$\Omega\I{P} = 20 \eV$,
the dependence of~$\check d(k\I{P})$ on~$k\I{P}$ is weak.
Additionally, quantities such as the line shape function~(\ref{eq:lineshape})
decrease rapidly as soon as~$\frac{\vec k\I{P}^2}{2}$ deviates
appreciably from~$\Omega\I{P}$.

Second, Auger decay is mediated by the rms two-electron matrix
element~(\ref{eq:AugME}).
To construct it, we need the direct two-electron matrix
element~$\acute \chi_h^{ij}(k\I{A}, \vartheta_{k\I{A}}, \varphi_{k\I{A}})$
[Eq.~(\ref{eq:twopwexpand})] for all~$h \in \Cal H$.
The rms matrix element in spherical polar coordinates is then
denoted by~$\tilde v(k\I{A}, \vartheta_{k\I{A}}, \varphi_{k\I{A}})$.
We examine three cases for the radial dependence~(\ref{eq:radialdep})
and (\ref{eq:radiusprod}).
This provides us with a good way to assess the quality of our
approximation.
First, following Refs.~\cite{Yakovlev:TR-02,Yakovlev:ES-03}, we use
\begin{equation}
  \label{eq:onemulti}
  \varrho_l(r, r') = \frac{r'^l}{r^{l+1}} \; .
\end{equation}
Second, we examine the reverse case,
\begin{equation}
  \label{eq:onemultiother}
  \varrho_l(r, r') = \frac{r^l}{r'^{l+1}} \; ,
\end{equation}
and, finally, the exact case~(\ref{eq:radialdep}).
For comparison with Refs.~\onlinecite{Yakovlev:TR-02,Smirnova:QC-03,%
Yakovlev:ES-03} and for computational efficiency, we will
use the crude approximation~(\ref{eq:onemulti}) throughout.

\begin{figure}
  \begin{center}
    \includegraphics[clip,width=\hsize]{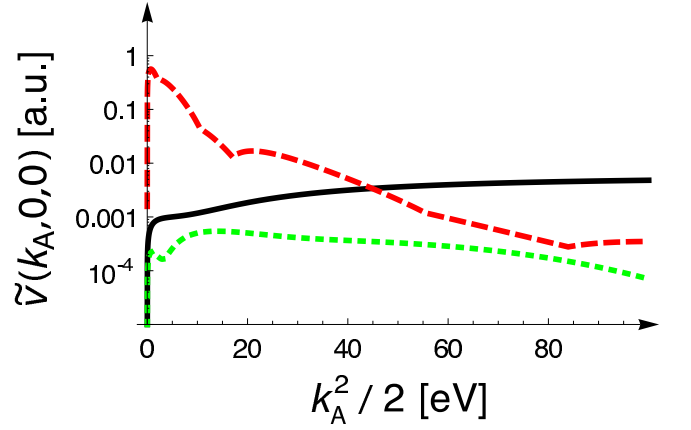}
    \caption{(Color online) The rms~two-electron direct matrix
             element for krypton~$3d$-hole Auger
             decay~$\tilde v(k\I{A}, 0, 0)$ of the
             essential-states model of Sec.~\ref{sec:essstmo}.
             Using the first radial dependence~(\ref{eq:onemulti}) to
             find~$\tilde v(k\I{A}, 0, 0)$ yields the solid black line;
             the second radial dependence~(\ref{eq:onemultiother}) gives the
             dashed red line;
             and the exact result~(\ref{eq:radialdep}) is represented by
             the dotted green line.}
    \label{fig:AugerMomWave}
  \end{center}
\end{figure}

We display the rms~two-electron matrix element~$\tilde
v(k\I{A}, 0, 0)$ in Fig.~\ref{fig:AugerMomWave} with
the viewing direction along the $z$~axis which, in turn, is the direction of
the linear \XUV{} and laser polarization vectors.
Our choice of direction agrees with Ref.~\onlinecite{Smirnova:QC-03}.
The dependence around the nominal Auger energy
of~$\Omega\I{A} = 40 \eV$---\ie, over the range plotted in
Fig.~\ref{fig:AugerOlga}---is weak.

\begin{figure}
  \begin{center}
    \includegraphics[clip,width=\hsize]{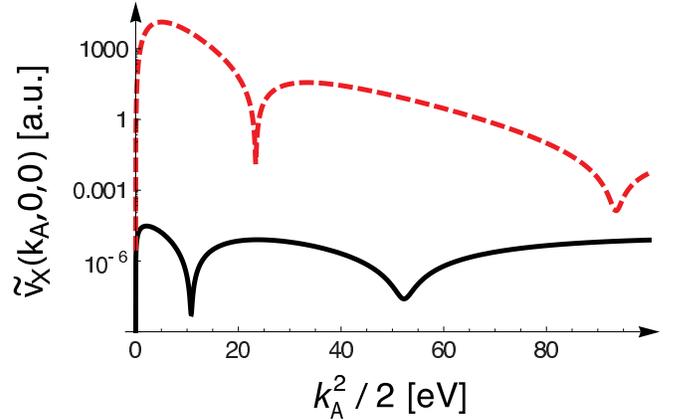}
    \caption{(Color online) The rms~two-electron exchange matrix
             element~$\tilde v\I{X}(k\I{A}, 0, 0)$ for krypton~$3d$-hole
             Auger decay.
             See Fig.~\ref{fig:AugerMomWave} for details.}
    \label{fig:AugerExchange}
  \end{center}
\end{figure}

To asses the impact of our omission of the two-electron exchange matrix
element~$v_{h \, \vec k\I{A} \, j \, i}$ in the essential-states model
of Sec.~\ref{sec:essstmo}, we compute its rms
value in spherical polar coordinates~$\tilde v\I{X}(k\I{A}, \vartheta_{k\I{A}},
\varphi_{k\I{A}})$.
It is displayed in Fig.~\ref{fig:AugerExchange} along the
$z$~axis, $\tilde v\I{X}(k\I{A}, 0, 0)$, for the
approximations in Eqs.~(\ref{eq:onemulti}) and (\ref{eq:onemultiother}).
The values of~$\tilde v\I{X}(k\I{A}, 0, 0)$ in our momentum range of interest
are roughly one order of magnitude smaller than
the corresponding values of the direct matrix element at the same
momentum in Fig.~\ref{fig:AugerMomWave} for the approximation
Eq.~(\ref{eq:onemulti}).
The other case [Eq.~(\ref{eq:onemultiother})] yields very large
value for~$\tilde v\I{X}(k\I{A}, 0, 0)$.
Particularly, these values are much larger than corresponding
values for~$\tilde v(k\I{A}, 0, 0)$ which is unphysical.
This comparison underscores that Eq.~(\ref{eq:onemulti}) represents a
reasonable approximation to the full Eq.~(\ref{eq:radialdep})
while Eq.~(\ref{eq:onemultiother}) does not.

%
%
%
\begin{table}
  \centering
  \begin{ruledtabular}
    \begin{tabular}{lccccc}
      $I\I{L,0}$ [$\U{\frac{W}{cm^2}}$] & $10^{10}$ & $10^{11}$ &
                              $10^{12}$ & $10^{13}$ & $10^{14}$ \\
      \hline
      $\Cal A\I{L}$ [a.u.] & 0.0094 & 0.030 & 0.094 & 0.30 & 0.94 \\
      $E\I{L,0}$ [a.u.]    & 0.00053 & 0.0017 & 0.0053 & 0.017 & 0.053 \\
      $U\I{P}$ [eV]        & 0.00060 & 0.0060 & 0.060 & 0.60 & 6.0 \\
      $\alpha\I{L}$ [$\angstrom$] & 0.087 & 0.28 & 0.87 & 2.8 & 8.7 \\
      $\frac{U\I{P}}{2 \, \omega\I{L}}$ & 0.00019 & 0.0019 &
                             0.019 & 0.19 & 1.9
    \end{tabular}
  \end{ruledtabular}
  \caption{Parameters for an $800 \U{nm}$~dressing laser.
           The vector potential amplitude~$\Cal A\I{L}$~(\ref{eq:lasVecPot}),
           the electric-field amplitude~$E\I{L,0}$~(\ref{eq:lasElField}),
           the ponderomotive potential~$U\I{P}$~(\ref{eq:pondpot}),
           the magnitude of the maximum
           excursion~$\alpha\I{L}$~(\ref{eq:maxexcurs}), and
           the second argument in generalized Bessel functions
           for a broad range of intensities~$I\I{L,0}$.}
  \label{tab:Upalpha}
\end{table}

We need to specify and characterize the \XUV{} and optical light
fields next.
A present-day attosecond-pulse light source typically has a
\XUV{}~peak intensity of at most~$I\I{X,0} = 10^{11} \U{\frac{W}{cm^2}}$
at a photon energy of~$\omega\I{X} = 90 \eV$.
The resulting ponderomotive potential~(\ref{eq:pondpot})
%
%
is~$1.8 \U{\mu eV}$ with a magnitude of maximum excursion~(\ref{eq:maxexcurs})
%
%
of~$8.2 \U{fm}$.
Clearly, the impact of the \XUV{}~field on photo- and Auger electrons
can be omitted in excellent approximation.
However, this approximation does not hold for the dressing
laser---typically a Ti:sapphire laser system
with near-infrared (\NIR{})~light of a wavelength of~$800 \U{nm}$ and
a photon energy of~$\omega\I{X} = 1.55 \eV$---which delivers a large range of
laser intensities.
Exemplary data are given in Table~\ref{tab:Upalpha}.
For lower intensities---up to~$10^{12} \U{\frac{W}{cm^2}}$---the
ponderomotive potential~(\ref{eq:pondpot}) and the vector potential
amplitude~$\Cal A\I{L}$ [Eq.~(\ref{eq:lasVecPot})] are small in
relation to the laser photon energy.
Also $\Cal A\I{L}$ is negligible compared with the momentum of the
photo- and Auger electrons.
Therefore, we may neglect the influence of the ponderomotive
potential~$U\I{P}$.
Specifically, this amounts to replacing generalized by ordinary Bessel
functions in equations such as~(\ref{eq:VolkovPhaseCW}), (\ref{eq:AugElSpec}),
(\ref{eq:xsectXUV}), and (\ref{eq:pelspec}).
For the Auger spectrum~(\ref{eq:AugElSpec}) this was done in
Refs.~\onlinecite{Yakovlev:TR-02,Yakovlev:ES-03}
and was done also in this paper for laser intensities up
to~$10^{12} \U{\frac{W}{cm^2}}$.
This approximation is assessed by comparing Auger spectra from
the expression using generalized Bessel functions~(\ref{eq:genBesDef})
with spectra from the corresponding expression using ordinary Bessel functions.
Excellent agreement is found.

\begin{figure}
  \begin{center}
    \includegraphics[clip,width=\hsize]{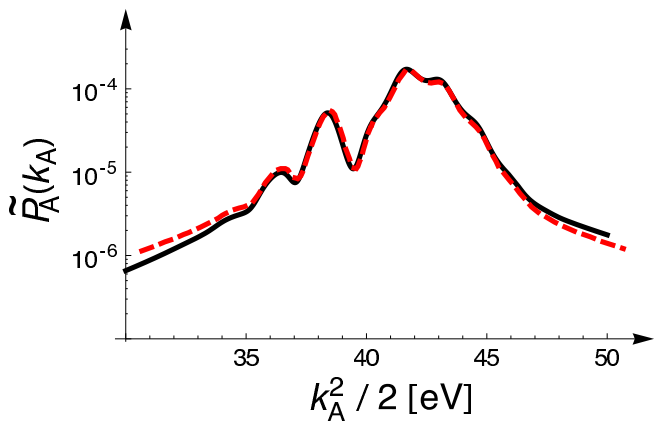}
    \caption{(Color online) Auger electron spectrum~(\ref{eq:AugElSpec})
             of laser-dressed krypton~$3d$-hole decay with an artificial width
             of~$\Gamma\I{broad} = 1.3 \eV$ for a dressing-laser
             intensity of~$5 \E{11} \U{\frac{W}{cm^2}}$.
             The solid black line was determined using our theory;
             the dashed red line is the solid curve of Fig.~3
             in Ref.~\onlinecite{Smirnova:QC-03} scaled by a factor
             of~$E\I{X,0}^2 = 2.9 \E{-4} \U{a.u.}$ and shifted by~$-0.68 \eV$.}
    \label{fig:AugerFig3}
  \end{center}
\end{figure}

Finally, we are in the position to put together all ingredients
to compute the laser-dressed Auger electron spectrum~(\ref{eq:AugElSpec})
in spherical polar coordinates of krypton~$3d$~hole
decay~\cite{footnote6}~$\acute{\Cal P}\I{A}(k\I{A},
\vartheta_{k\I{A}}, \varphi_{k\I{A}})$ integrated over the azimuth
angle~$\tilde{\Cal P}\I{A}(k\I{A})
= 2\pi \> \acute{\Cal P}\I{A}(k\I{A}, 0, 0)$.
As a verification of our solution, we compute the
Auger spectrum in Fig.~\ref{fig:AugerFig3} for a dressing-laser
intensity of~$5 \E{11} \U{\frac{W}{cm^2}}$
as in Fig.~3 in Ref.~\onlinecite{Smirnova:QC-03}.
Our result agrees very well with the one in Ref.~\cite{Smirnova:QC-03} apart
from an overall scaling factor
%
%
of~$E\I{X,0}^2 = 2.9 \E{-4} \U{a.u.}$ due to the \XUV{}~electric-field
strength which was set to unity in Ref.~\onlinecite{Smirnova:QC-03}.
Finally, the spectrum in Fig.~3 of Ref.~\onlinecite{Smirnova:QC-03}
was determined with a nonzero value for~$\Delta\I{R}$.
We find that we need to shift it by a value of~$-0.68 \eV$
[Sec.~\ref{sec:compdet}] to achieve agreement.
The structure in the figure is mostly due to the time dependence as the
momentum dependence of the dipole and Auger decay matrix elements
is weak.
The differences between both curves are ascribed to a somewhat different
treatment of the matrix elements.

\begin{figure}
  \begin{center}
    \includegraphics[clip,width=\hsize]{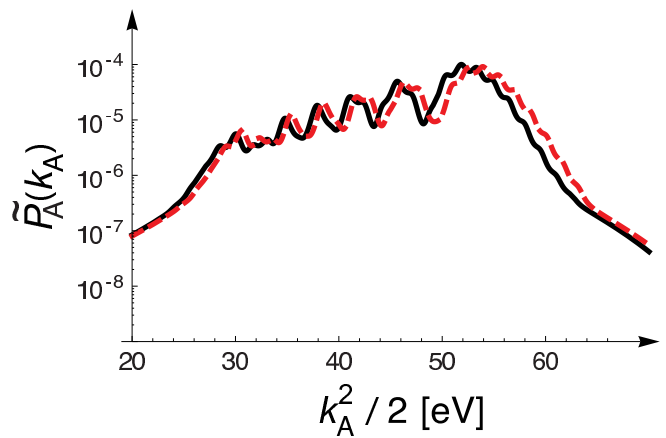}
    \caption{(Color online) Auger electron spectrum~(\ref{eq:AugElSpec})
             of laser-dressed krypton~$3d$-hole decay for an artificial width
             of~$\Gamma\I{broad} = 1.3 \eV$ and a dressing-laser
             intensity of~$10^{13} \U{\frac{W}{cm^2}}$.
             The solid black line was determined using generalized Bessel
             functions~(\ref{eq:genBesDef});
             the dashed red line was obtained neglecting~$U\I{P}$
             and using ordinary Bessel functions.}
    \label{fig:AugerOlga}
  \end{center}
\end{figure}

We present the Auger electron spectrum for a dressing-laser intensity
of~$10^{13} \U{\frac{W}{cm^2}}$ in Fig.~\ref{fig:AugerOlga}~\cite{footnote7}.
The spectrum is first computed using generalized Bessel functions
and a nonvanishing ponderomotive potential in Eq.~(\ref{eq:AugElSpec}).
Then, it is determined using ordinary Bessel functions
and~$U\I{P} = 0$.
There are clear differences between the spectra.
The convergence with respect to the number of terms in
Eq.~(\ref{eq:genBesDef}) is rapid;
a summation from~$-2$ to~$+2$ in Eq.~(\ref{eq:genBesDef}) was sufficient
We conclude that dressing-laser intensities around and
above~$10^{13} \U{\frac{W}{cm^2}}$ require an accurate treatment
of the Volkov phase~(\ref{eq:VolkovExpand}).
We need 15~laser photon indices (Bessel functions) to account for
absorption and emission of laser photons in the sum~(\ref{eq:lasdreauampre})
for the Auger electron spectrum.
This is in good agreement with previous studies of $K$-shell ionization of
laser-dressed neon~\cite{Buth:ET-07}, argon~\cite{Buth:AR-08},
and krypton~\cite{Buth:TX-07} atoms---however, with a very
different theoretical approach---where 20, 12, and 5~photon blocks
were required, respectively, to converge the calculations for the
same dressing laser parameters that are used here.
Seemingly crucial for the necessary number of photon blocks are the
decay widths of the inner-shell hole which were~$0.27 \eV$, $0.66 \eV$, and
$2.7 \eV$, respectively.
Our artificial value for the krypton~$3d$ decay width of~$1.3 \eV$ lies
between the decay widths of the krypton and the argon $K$~shell vacancies.

\section{Conclusion}
\label{sec:conclusion}

We have devised and applied an \emph{ab initio} theory for inner-shell
\XUV{}~photoionization and subsequent Auger decay
of laser-dressed atoms, a so-called two-color problem.
Our work aims at the study and control of electron correlations---here
manifested in terms of electronic decay---which is the most profound
goal of attosecond science.
The photo- and Auger electrons experienced an optical dressing laser
which was considered to be intense but not strong enough to excite
or ionize electrons in the atomic ground state.
We used the Hartree-Fock-Slater~(HFS)~approximation as a starting point for
the description of the atomic electronic structure.
The HFS~orbitals were then used to represent the full Hamiltonian.
We employed a single configuration-state function to represent the ground
state and singly- and doubly-excited states.
The light fields were treated semiclassically and we use the strong-field
approximation to treat the influence of the optical laser on the
photo- and Auger electrons.
The influence of the laser on the atomic ground-state electrons
was neglected.
The quantum dynamics of the problem was described in terms of
equations of motion~(EOMs).
The EOMs were solved analytically for an essential-states model and a
closed-form expression for the Auger electron amplitude was obtained.
Furthermore, the \XUV{}-absorption cross section of laser-dressed atoms
and an expression for the laser-dressed photoelectron spectrum were derived.
We applied our formalism to study the photoionization of a $3d$~orbital
($M$~shell) of a krypton atom and its subsequent $M_{4,5} N_1 N_{2,3}$~Auger
decay where the vacancy is filled with a $4s$~valence electron expelling a
$4p$~valence electron.
Following Ref.~\onlinecite{Smirnova:QC-03}, we assumed an artificial
decay width of krypton~$3d$~vacancies of~$1.3 \eV$.
The atomic orbitals were approximated by suitably scaled hydrogen wave
functions circumventing the need for a HFS~computation.
We discussed the approximations made and studied the convergence of the
Auger decay matrix element and the expansion in terms of generalized
Bessel functions.
We compared our laser-dressed Auger spectrum to literature results
of Smirnova~\etal~\cite{Smirnova:QC-03} and found good agreement.
Finally, we presented the Auger electron spectrum
for~$10^{13} \U{\frac{W}{cm^2}}$~\NIR{}~laser intensity.

Our work opens up a multitude of future research perspectives.
We have devised a general \emph{ab initio} framework which allows us
to create simplified models of varying sophistication tailored
to model many physical situations in laser-dressed Auger decay.
In this paper, we reduced our EOMs to an essential-states model
which can be solved analytically and comprises sufficient details for
a number of physical problems.
In a forthcoming paper~\cite{Buth:AD-09}, we will use it to investigate
coherence and interference of Auger electrons and their control by a laser.
In a next step, our model can be generalized to a few
states, \eg, all states in a subshell can be considered, avoiding
magnetic quantum number-averaged dipole and Auger matrix elements.

\begin{acknowledgments}
We thank Olga Smirnova and Stephen H.{} Southworth for fruitful discussions.
C.B.~was supported by the National Science Foundation under
Grant~No.~PHY-0701372.
\end{acknowledgments}


\begin{thebibliography}{68}
\bibitem[{foo({\natexlab{a}})}]{footnote1}
\bibinfo{note}{We use the form of a Volkov wave~\cite{Madsen:SF-05} which is
  the solution of the time-independent Schr\"odinger equation for the
  interaction with the electric field in velocity
  form~(\ref{eq:FreeLaserHam})---the interaction with the \XUV{}~light is in
  length form~(\ref{eq:xuvint})---instead of the length form which is used in
  Ref.~\onlinecite{Smirnova:QC-03}. The difference of Eq.~(4) in
  Ref.~\onlinecite{Smirnova:QC-03} to our Eq.~(\ref{eq:VolkovState}) is the
  missing factor~$\euler^{\imag \, \vec A\I{L}(t) \mul \vec r}$ in our
  expression~\cite{Scully:QO-97}.}
\bibitem[{foo({\natexlab{b}})}]{footnote2}
\bibinfo{note}{Our expressions are written for \XUV{}~light only; yet, in the
  corresponding expressions in Ref.~\onlinecite{Smirnova:QC-03}, laser dressing
  is also considered. The slight deviations between both sets of equations due
  to these different situations are obvious and not mentioned here.
  Equation~(10) in Ref.~\onlinecite{Smirnova:QC-03} differs from our
  Eq.~(\ref{eq:EOMxuvOlgaFin}) by a factor of~$\frac{1}{\sqrt{2}}$ in the first
  term on the right-hand side and by a factor of~$-2 \, \sqrt{2}$ in the second
  term. Further, Eq.~(11) in Ref.~\onlinecite{Smirnova:QC-03} agrees with our
  Eq.~(\ref{eq:EOMaugerOlgaFin}) apart from a factor of~$\frac{-1}{\sqrt{2}}$.}
\bibitem[{foo({\natexlab{c}})}]{footnote3}
\bibinfo{note}{The difference between our expression~(\ref{eq:EOMholeAmp})---in
  the limit of vanishing laser intensity---and Eq.~(15) in
  Ref.~\onlinecite{Smirnova:QC-03} is a factor~$\frac{-1}{\sqrt{2}}$ due to our
  more elaborate treatment of many-electron effects. Further the minus sign
  before~$\Delta$ in Eq.~(15) in Ref.~\onlinecite{Smirnova:QC-03} is a
  misprint. Smirnova~\etal{}~\cite{Smirnova:QC-03} found that their Eq.~(15)
  without laser field is an excellent approximation for weak laser intensities
  to a numerically exact solution of the coupled differential equations with a
  laser field.}
\bibitem[{foo({\natexlab{d}})}]{footnote5}
\bibinfo{note}{There is no additional factor of~2 in Eq.~(\ref{eq:xsect})
  compared with Eq.~(31) in Ref.~\onlinecite{Buth:TX-07} to account for the
  number of electrons in a spatial orbital because our determinantial approach
  treats the transition from spin orbitals to spatial orbitals fully.}
\bibitem[{foo({\natexlab{e}})}]{footnote4}
\bibinfo{note}{The energy of the krypton~$3d$~shell of~$-96.6 \eV$ which is
  used in Ref.~\onlinecite{Yeh:AS-85} is lower than the value of~$-70 \eV$
  assumed for~$\varepsilon_{3d}$ in this paper. The photon energy needs to be
  adjusted accordingly to~$116.6 \eV$ to produce photoelectrons of the same
  kinetic energy in both cases.}
\bibitem[{foo({\natexlab{f}})}]{footnote6}
\bibinfo{note}{In the experiments on Auger decay of krypton
  $3d$~holes~\cite{Drescher:TR-02}, the photoelectron was not observed and thus
  we integrate over its momentum [see Eq.~(\ref{eq:AugElSpec})]. In doing so,
  we neglect the impact of the laser on the photoelectrons. This is justified
  by considering Ref.~\cite{Buth:TX-07}, from which we know that the area under
  the cross section with and without laser dressing is approximately conserved.
  Hence, in our quite similar situation, we can assume that the integration
  over the photoelectron momentum does not vary noticeably whether the laser
  dressing of the photoelectrons is included or not.}
\bibitem[{foo({\natexlab{g}})}]{footnote7}
\bibinfo{note}{The Auger spectrum in Fig.~\ref{fig:AugerOlga} is very broad and
  extends down to~$\sim 25 \eV$. As the nominal photoelectron energy is~$20
  \eV$ and the photoelectron spectrum is similarly broadened, the photo- and
  the Auger electron spectra overlap. To avoid this, a higher \XUV{}~photon
  energy may be used.}
\end{thebibliography}
\end{document}